\documentclass[%
reprint,
superscriptaddress,
 amsmath,amssymb,
 aps,
]{revtex4-2}

\providecommand*{\code}[1]{\texttt{#1}}
\usepackage{babel}
\usepackage{array}
\usepackage{float}
\usepackage{multirow}
\usepackage{amsmath}
\usepackage{amssymb}
\usepackage{cancel}
\usepackage{stackrel}
\usepackage{graphicx}
\usepackage{lmodern}
\usepackage{listings}
\usepackage{multirow}
\usepackage{float}
\usepackage{graphicx}
\usepackage{dcolumn}
\usepackage{bm}
\usepackage{bbm}
\usepackage{siunitx}
\usepackage{hyperref}

\begin{document}
\title{Reproducibility and control of superconducting flux qubits}

\author{T. \surname{Chang}}
\affiliation{Quantum Nanoelectronics Laboratory, Department of Physics \& Bar-Ilan
Institute of Nanotechnology and Advanced Materials (BINA), 5290002
Ramat-Gan, Israel.}

\author{I. \surname{Holzman}}
\affiliation{Quantum Nanoelectronics Laboratory, Department of Physics \& Bar-Ilan
Institute of Nanotechnology and Advanced Materials (BINA), 5290002
Ramat-Gan, Israel.}

\author{T. \surname {Cohen}}
\affiliation{Quantum Nanoelectronics Laboratory, Department of Physics \& Bar-Ilan
Institute of Nanotechnology and Advanced Materials (BINA), 5290002
Ramat-Gan, Israel.}

\author{B. C. \surname{Johnson}}
\affiliation{ARC Centre for Quantum Computation and Communication Technology
(CQC2T) \& School of Physics, University of Melbourne, Parkville,
3010, VIC, Australia.}

\author{D. N. \surname{Jamieson}}
\affiliation{ARC Centre for Quantum Computation and Communication Technology
(CQC2T) \& School of Physics, University of Melbourne, Parkville,
3010, VIC, Australia.}

\author{M. \surname{Stern}}
\affiliation{Quantum Nanoelectronics Laboratory, Department of Physics \& Bar-Ilan
Institute of Nanotechnology and Advanced Materials (BINA), 5290002
Ramat-Gan, Israel.}

\date{15 June 2022}

\newcommand{\hamil}{\mathcal{H}}
\newcommand{\phir}{\Phi_{R}}
\newcommand{\phis}{\Phi_{S}}
\newcommand{\phisr}{\Phi_{S/R}}
\newcommand{\sorr}{S\left(R\right)}
\newcommand{\ixs}{I_{x, S}}
\newcommand{\izs}{I_{z, S}}
\newcommand{\ixr}{I_{x, R}}
\newcommand{\izr}{I_{z, R}}
\newcommand{\ixsr}{I_{x, S/R}}
\newcommand{\izsr}{I_{z, S/R}}
\newcommand{\gfsr}{\Gamma_{S/R}}
\newcommand{\gfs}{\Gamma_{S}}
\newcommand{\gfr}{\Gamma_{R}}
\newcommand{\gfso}{\Gamma_{2nd}}
\newcommand{\gffo}{\Gamma_{1st}}
\newcommand{\micron}{\si{\micro\meter}}
\newcommand{\br}[1]{\Bra\{#1\}}
\newcommand{\kt}[1]{\Ket\{#1\}}
\newcommand{\brkt}[1]{\Braket\{#1\}}

\newcommand{\microsec}{\si{\micro\second}}


\begin{abstract}
Superconducting flux qubits are promising candidates for the physical
realization of a scalable quantum processor. Indeed, these circuits
may have both a small decoherence rate and a large anharmonicity.
These properties enable the application of fast quantum gates with
high fidelity and reduce scaling limitations due to frequency crowding.
The major difficulty of flux qubits' design consists of controlling
precisely their transition energy - the so-called qubit gap - while
keeping long and reproducible relaxation times. Solving this problem
is challenging and requires extremely good control of e-beam lithography,
oxidation parameters of the junctions and sample surface. Here we
present measurements of a large batch of flux qubits and demonstrate
a high level of reproducibility and  control of 
qubit gaps ($\pm\SI{0.6}{\giga\hertz}$), 
relaxation times ($15-\SI{20}{\micro\second}$) and 
pure echo dephasing times ($15-\SI{30}{\micro\second}$). 
These results open the way for potential applications in the
fields of quantum hybrid circuits and quantum computation. 
\end{abstract}

\maketitle

Thanks to their long coherence times and ease of use \citep{PhysRevLett.107.240501,Place2021,PhysRevLett.111.080502},
transmon qubits are today one of the most popular architectures for
building superconducting quantum processors \citep{Arute2019}. Yet,
as one scales up the system, the large eigenvalue manifold of each
transmon generates issues related to frequency crowding and gate fidelity
\citep{PhysRevApplied.10.034050}. In contrast to transmons, superconducting
flux qubits \citep{Mooij1999,PhysRevB.60.15398,vanderWal2000,Chiorescu2003}
intrinsically possess a huge anharmonicity: the higher energy levels
of the system are very far from the qubit transition. Consequently,
the flux qubit behaves as a \emph{true} two level system, which limits
frequency crowding problems. Moreover, it can be manipulated on a
much shorter timescale ($<\SI{10}{\nano\second}$) and therefore could
potentially exhibit better gate fidelity. In addition, this architecture
offers interesting prospects for the development of hybrid quantum
circuits since its large magnetic dipole could allow for an efficient
transfer of quantum information between isolated quantum systems,
such as spins in semiconductors \citep{PhysRevLett.105.210501,PhysRevB.81.241202,PhysRevA.92.052335}.

The two major issues of flux qubit designs are device-to-device gap
reproducibility and coherence \citep{PhysRevLett.95.257002,PhysRevLett.97.167001,PhysRevLett.105.237001,Bylander2011}.
The flux qubit transition energy - the so-called qubit gap- is difficult
to control and requires an extremely precise tuning of the fabrication
parameters. Moreover, the flux qubit coherence times are known for
their sizeable irreproducibility. Long coherence times reported in
previous works relate only to a few singular flux qubits \citep{Bylander2011}.
In the last years, flux qubits embedded in 3D cavities \citep{PhysRevLett.113.123601}
or in coplanar resonators \citep{PhysRevB.93.104518} have exhibited
more reproducible and generally improved relaxation times. More recently,
a new design - the so-called capacitively shunt flux qubit - has shown
even better coherence times \citep{Yan2016}. However, this same shunting
capacitance used to better control the qubit strongly decreases its
anharmonicity to a level which becomes almost comparable to that of
a transmon. Clearly, further improvements in coherence times and in
control are necessary if the flux qubit is to be an alternative option
for quantum computation.

\begin{figure*}[t]
\includegraphics[width=1\textwidth]{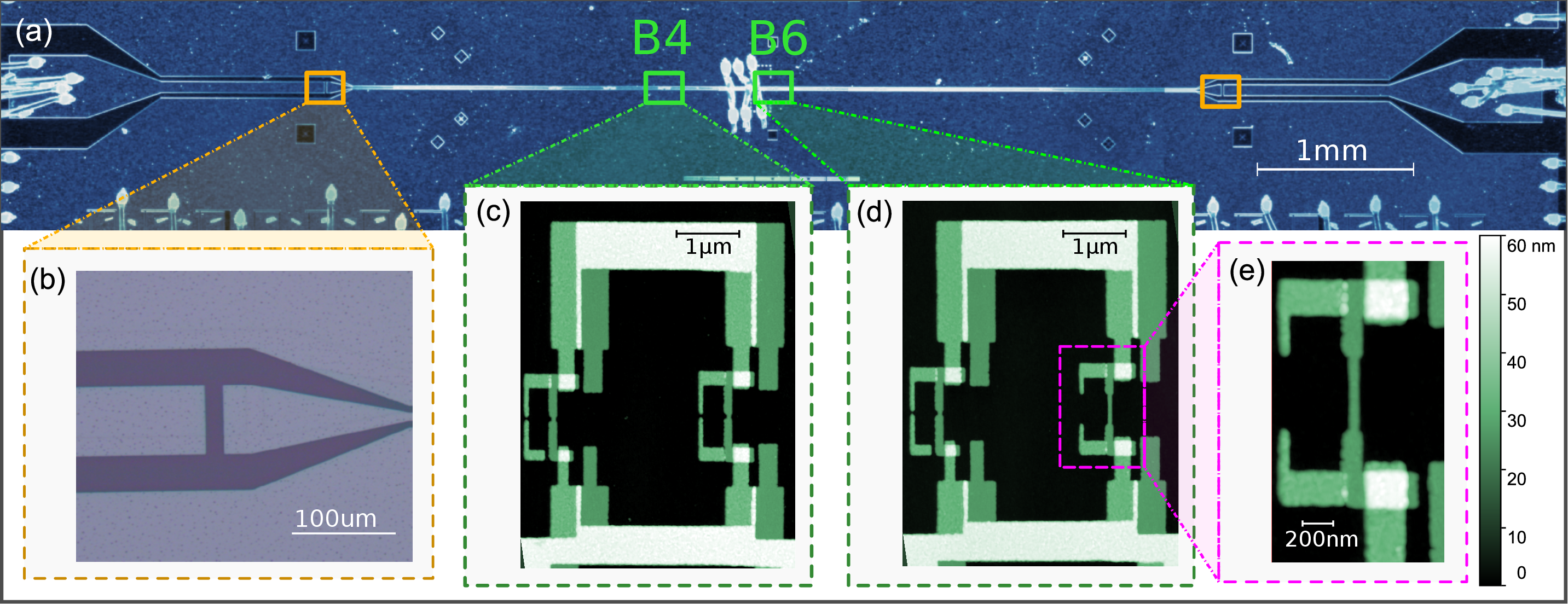}\caption{\textbf{Circuit implementation.} (a) Optical microscope image of a
$\lambda/2$ CPW resonator (resonator B) intersected and galvanically
coupled to a series of eleven flux qubits labelled $B1$ to $B11$.
The CPW resonator length is chosen to be 5.73 mm, such that the first
resonant mode is at $f_{rB}\simeq9.8$ GHz. (b) Close up view of the
coupling capacitor terminating at both ends the CPW resonator. The
value of the capacitance is calculated by an electromagnetic simulator
(Sonnet) to be $C_{C}\sim\SI{5.0}{\femto\farad}$. (c) Colored AFM
micrograph of qubit $B4$. The surface area of the unitary junction
is $A_{uni}=0.0526\pm\SI{0.0008}{\micro\meter}^{2}${\scriptsize{}{}
}and the small junction was chosen to have $\alpha=0.5$. (d) Colored
AFM micrograph of qubit $B6$. The surface area of the unitary junction
and the ratio $\alpha$ are identical to $B4$. The loop of this qubit
includes a thin constriction. (e) Close up view of the 30-nm width
constriction of qubit $B6$.\label{fig:circ-impl}}
\end{figure*}

In this work, we present a good improvement in the control and reproducibility
of these qubits. We present a systematical study of a large batch
of more than twenty devices and demonstrate that it is possible to
control their gap energy to within less than $\SI{1}{\giga\hertz}$
while obtaining reproducible relaxation times $T_{1}\sim15-\SI{20}{\micro\second}$
and pure dephasing times $T_{2E}^{\phi}\sim 15-\SI{30}{\micro\second}$.
This reproducibility enabled us to analyze the different factors that
impede the coherence times and systematically eliminate them. Our
work opens new perspectives for potential applications in the fields
of quantum hybrid circuits and quantum computation.


Our method explores the role of the substrate in device variability
by employing a standard gate oxide process based on other applications
of CMOS device technology \citep{Pla2012}. The three samples presented
in this work are fabricated on silicon chips and contain a 150-nm
thick aluminium coplanar waveguide (CPW) resonator, with two symmetric
ports used for microwave transmission measurements (see Figure \ref{fig:circ-impl}(a)).
The CPW resonator A is directly fabricated on a high resistivity ($>\SI{10}{\kilo\ohm\centi\meter}$)
silicon wafer with native oxide while resonators B and C are fabricated
on a 5 nm thermally grown silicon oxide layer. A series of eleven
flux qubits is galvanically coupled to each CPW resonator. In the
following, the qubits are labelled according to their spatial position
on the relevant resonator (e.g. $A1...A11,B1...B11,C1...C11)$.

Our flux qubit design consists of a superconducting loop intersected
by four Josephson junctions, one of which is smaller than the others
by a factor $\alpha$. This circuit behaves as a two-level system
when the flux threading the loop is close to half a flux quantum $\Phi\sim\Phi_{0}/2$
\citep{Mooij1999,PhysRevB.60.15398}. Each level is characterized
by the direction of a macroscopic persistent current $I_{P}$ flowing
in the loop of the qubit. The value of the persistent current $I_{P}$
- typically of the order of 200-$\SI{300}{\nano\ampere}$ - gives rise to a huge magnetic
moment ($\sim500\,\mathrm{GHz/G})$, making the energy of each level
very sensitive to external magnetic flux. At $\Phi=\Phi_{0}/2$, the
two levels are degenerate, hybridise and give rise to an energy splitting
$h\Delta$ called the flux-qubit gap. At this point, the qubit is
immune to flux noise at first order and should exhibit a long coherence
time.

Figure \ref{fig:circ-impl}(c) and (d) present Atomic Force Microscope
(AFM) images of qubits $B4$ and $B6$. The loop area of qubit $B4$
(resp. $B6$) is $S_{B4}\simeq\SI{13.3}{\micro\meter}^{2}$ (resp.
$S_{B6}\simeq\SI{12.4}{\micro\meter}^{2}$). The three identical junctions
have a Josephson energy $E_{J}/h=\SI{360}{\giga\hertz}$ and a single
electron charging energy $E_{C}/h=\SI{3.68}{\giga\hertz}$ while the
fourth junction is smaller than others by $\alpha$=0.5. In addition,
qubit $B6$ contains a 30 nm width constriction over a length of 500
nm (see Figure \ref{fig:circ-impl}(e)). The qubits are fabricated
by e-beam lithography with a tri-layer CSAR-Ge-MAA process (See \citep{sup}
for more details). The germanium mask is rigid and robust to the oxygen
ashing cleaning step. Moreover, it dissipates efficiently the charges
during e-beam lithography and thus provides an excellent precision
and reproducibility of the junction sizes. The electron-beam lithography
is followed by double angle-evaporation of Al--AlOx--Al performed
at a well controlled temperature ($-\SI{50}{\celsius}/+\SI{7}{\celsius})$. The low
temperature enables us to reduce the grain size of aluminium, to better
control the dimensions and oxidation of our junctions and to fabricate
small constrictions with high fidelity.

\begin{figure*}
\includegraphics[width=1\textwidth]{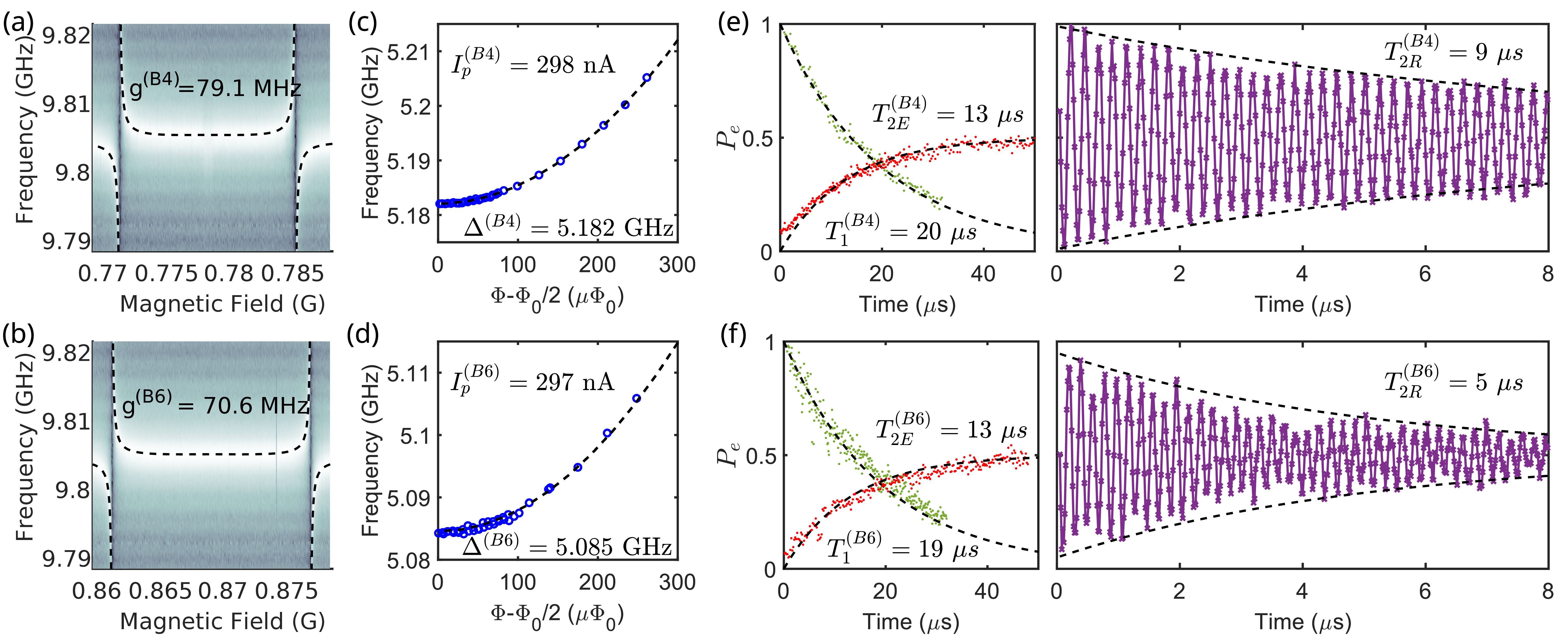}\caption{\textbf{Characterisation of qubit $B4$ (top panels) and $B6$ (bottom
panels).} (a-b) Transmission spectrum of CPW resonator B versus applied
magnetic field showing anticrossing of qubit B4/B6. For each qubit
we fit the anticrossing to our qubit-resonator coupling model and
extract the value of the coupling constant $g$ of the qubit with
the resonator. (c-d) Measured qubit frequency (blue dots) and fit
(black dashed curve) yielding the qubit parameters $\Delta$ and $I_{p}$.
(e-f) (left panel) Qubit energy relaxation and spin echo measurements.
The excited state probability $P_{e}$ is plotted as a function of
the delay between the $\pi$ pulse and the readout pulse (green dots)
or between the two $\pi/2$ pulses of the echo sequence (red dots).
The black dashed line is an exponential fit to the energy relaxation
(spin-echo) data. (Right panels) Measured Ramsey fringes (purple solid
line) with fit to its exponentially decaying envelope.\label{fig:charac-qb-B4B6}}
\end{figure*}

We first characterize the qubit-resonator system by spectroscopic measurements (see \citep{sup} for experimental setup). Figure \ref{fig:charac-qb-B4B6}(a-b) shows a continuous wave transmission scan of resonator
$B$ taken as a function of the applied magnetic field. This measurement
is performed with a vanishing power corresponding to an average of
less than one photon in the resonator. We observe an anticrossing
each time a qubit and the resonator are resonant. Far from the anticrossings,
the resonance corresponding to the first mode of the resonator is
$f_{rB}=9.804$ GHz and its quality factor is $Q_{B}=2800$ \citep{sup}.

The frequency dependence of qubit $B4$ and $B6$ on $\Phi$ is shown
in Figure \ref{fig:charac-qb-B4B6}(c-d), respectively. The transition
frequency of each qubit follows $f_{01}=\sqrt{\Delta^{2}+\varepsilon^{2}}$
with $\varepsilon=2I_{P}\left(\Phi-\Phi_{0}/2\right)/h$, yielding
$\Delta^{(B4)}=\SI{5.182}{\giga\hertz}$ and $I_{P}^{(B4)}=\SI{298}{\nano \ampere}$  (resp. $\Delta^{(B6)}=\SI{5.085}{\giga \hertz}$, $I_{P}^{(B6)}=\SI{297}{\nano\ampere}$). Since both qubits were designed to have
the same parameters, this demonstrates excellent reproducibility of
our e-beam lithography and oxidation parameters. Taking into account
the contribution of geometric capacitance between neighboring islands
allows us to fit the parameters of the flux-qubits in good agreement
with the measured values of $\alpha$ and $E_{J}$ extracted from
Ambegaokar-Baratoff formula (see \citep{sup}). We now turn to the
coherence times at the so-called optimal point where the qubit frequency
$f_{01}=\Delta$ is insensitive to first order to flux-noise \citep{PhysRevLett.95.257002,PhysRevLett.97.167001}.
Energy relaxation decay is shown in Figure \ref{fig:charac-qb-B4B6}(e)
and f to be exponential for both qubits, with $T_{1}=\SI{20}{\micro\second}$
for $B4$ and $\SI{19}{\micro\second}$ for $B6$. Ramsey fringes
show an exponential decay for $B4$ with $T_{2R}=\SI{9}{\micro\second}$,
for $B6$ with $T_{2R}=\SI{5}{\micro\second}$. Spin-echo decays exponentially
with identical dephasing times $T_{2E}=\SI{13}{\micro\second}$. Apparently,
the presence of the constriction in qubit $B6$ does not seem to influence
the coherence time of the qubit. This property is particularly exciting
if one wishes to coherently couple a single spin to this circuit \citep{PhysRevA.92.052335}.

\begin{figure*}
\includegraphics[width=1\textwidth]{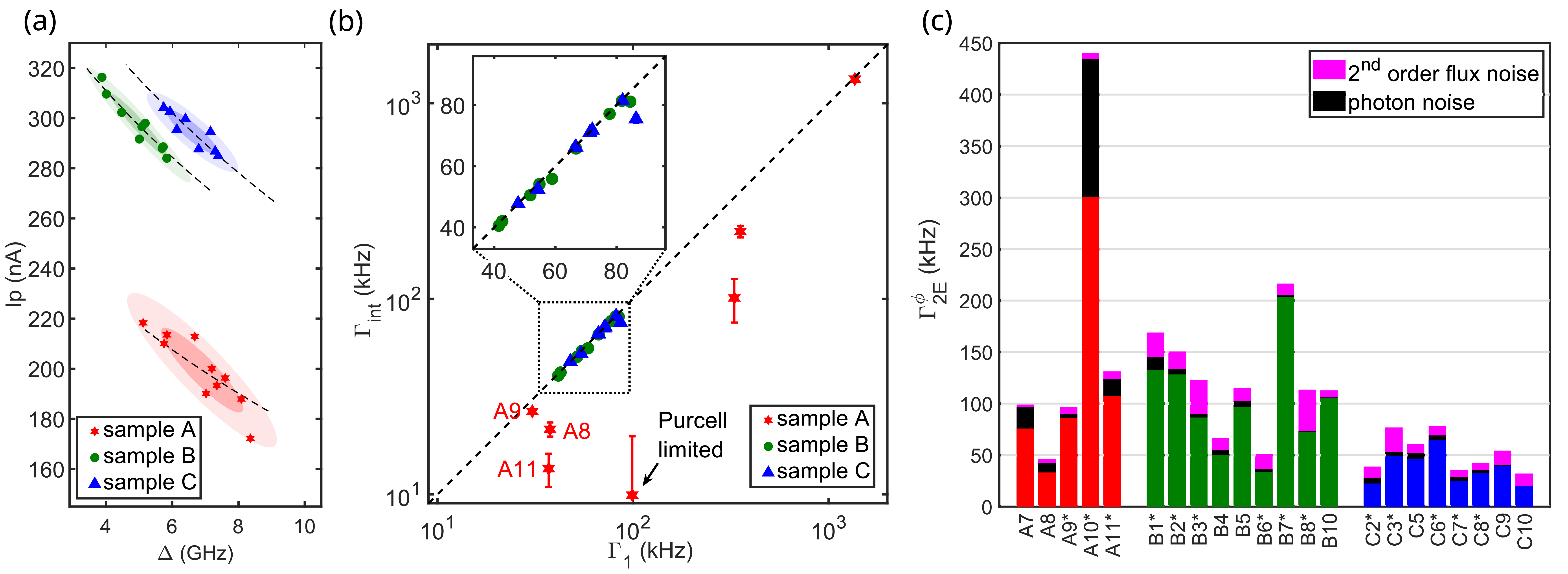}\caption{\textbf{Reproducibility and control.} (a) Persistent current $I_{P}$
versus gap $\Delta$ of the qubits of sample A (native oxide, red
stars), sample B ($\SI{5}{\nano\meter}$ grown silicon oxide layer,
green dots) and sample C ($\SI{5}{\nano\meter}$ grown silicon oxide
layer, blue triangles). The colored regions are obtained by assuming
a normal distribution along axes defined by principal component analysis
(PCA). The probability to find a qubit within the dark (resp. light)
colored area is 50\% (resp. 90\%). The dashed black lines are obtained
by numerical simulations of the flux qubits (see \citep{sup}) at
their average value $\left(\left\langle E_{J}\right\rangle _{A/B/C},\left\langle E_{c}\right\rangle _{A/B/C},\left\langle \alpha\right\rangle _{A/B/C}\right)$
while varying the parameter $\alpha$ by $\pm5\%$. (b)  The estimated intrinsic relaxation rates $\Gamma_{int}=\Gamma_1-\Gamma_P$ versus measured  relaxation rates $\Gamma_{1}$ for qubits of sample A (red stars), B (green dots) and C (blue triangles). The error bars stem from calibration uncertainties of $\pm\SI{1}{\decibel}$  of the incoming power at the resonator input. (c) Stacked bar chart showing the pure dephasing rates
$\Gamma_{2E}^{\phi}$ at optimal points of the measured qubits of
sample A (red), B (green) and C (blue). The black color corresponds to the calculated decoherence rate due to photon noise in the resonator. The pink color corresponds to calculated decoherence rate due to second order flux noise. The black stars indicate the presence of a 30-nm width constriction in the loop
of the qubit. The presence of a constriction does not seem to affect significantly 
the relaxation or the dephasing of the qubits.\label{fig:rep-control-t1-t2}}
\end{figure*}

We repeat this procedure for the qubits of our three samples. Each
qubit is thus characterized by its spectroscopic parameters $\Delta$
and $I_{P}$, extracted from the dependence of its transition frequency
on the applied flux. In Figure \ref{fig:rep-control-t1-t2}(a), we
represent a graph showing the gaps $\Delta$ of the different qubits
versus their persistent currents $I_{P}$.  In order to optimize our qubit design, we varied the size of the unitary junctions of samples A, B and C while keeping an approximately constant critical current density of $\sim\SI{13.5}{\micro\ampere\per\micro\meter\squared}$. Within each sample, the
qubit parameters $\left(E_{J},E_{C},\alpha\right)$   were designed
to be identical and thus the qubits should be clustered within a well
defined region. The extent of this region indicates the level of reproducibility
of our fabrication process. A slight improvement in the data spread
is observed for Sample B and C in comparison to sample A. Quantitatively
speaking, the gap average values are $6.9\pm1$GHz, $5.1\pm0.7$ GHz
and $6.6\pm0.6$ GHz for samples A, B and C respectively. A principal
component analysis (PCA) is performed on the covariance matrix of
the $(\Delta,I_{P})$ data-points in order to define regions with
high probability to find a qubit. For each sample, a dashed line is
represented and corresponds to the result of qubit numerical diagonalizations
(see \citep{sup}) while varying the parameter $\alpha$ by $\pm5\%$
around their respective average value $\left(\left\langle E_{J}\right\rangle _{A/B/C},\left\langle E_{c}\right\rangle _{A/B/C},\left\langle \alpha\right\rangle _{A/B/C}\right)$.
For the three samples, the principal axis and the numerical diagonalizations
are well aligned indicating that the main origin of disorder is indeed
uncontrolled variations of the value of the parameter $\alpha$. 
The variation of the critical current density of the junctions due to different oxidation of  samples A, B and C ($\pm5\% $) leads to an additional uncertainty of $\pm\SI{150}{\mega\hertz}$ in the
control of the desired qubit gap.

In Figure \ref{fig:rep-control-t1-t2}(b), we represent the spread of the relaxation rates  $\Gamma_{1}$ of the different qubits. Qubit $A9$ exhibits the longest relaxation time with $T_1=\SI{32}{\micro\second}$. Several mechanisms contribute to relaxation of qubits; among them, spontaneous
emission by the qubit to the resonator (the so-called Purcell effect
\citep{PhysRevLett.101.080502}). The Purcell rate $\Gamma_{P}$ can
be quantitatively determined by measuring the qubit Rabi frequency
$\Omega_{R}$ for a given microwave power $P_{in}$ at the resonator
input. For a qubit coupled symmetrically to the input and output lines,
a simple expression for  $\Gamma_{P}$ was obtained in Ref. \citep{PhysRevLett.113.123601}. We thus calculated
$\Gamma_{P}$ for each qubit and represented the intrinsic relaxation
rates of the qubits defined as $\Gamma_{int}=\Gamma_{1}-\Gamma_{P}$. The average values
of the intrinsic relaxation rates are $260\pm440$ kHz, $61\pm15$
kHz and $68\pm11$ kHz for samples A, B and C, respectively. These
average numbers are comparable to those obtained in Ref. \citep{Yan2016}
for C-shunted flux qubits. Relaxation due to $1/f^{\gamma}$-flux noise can be safely neglected
for qubits in our frequency range \citep{Yan2016}.  The spread of the relaxation rates in sample
B and C is remarkable compared to sample A and more generally to the state of the art \citep{PhysRevB.93.104518,PhysRevLett.113.123601}. We thus come to the conclusion that better
qubit reproducibility in terms of relaxation rates is obtained on
samples with a thermally grown 5 nm width silicon oxide layer. It
is yet important to stress that the best relaxation rates ($\sim\SI{25}{\kilo\hertz}$)
were obtained on intrinsic silicon (e.g. A11, A8). These findings
are consistent with previous studies comparing loss tangents for silicon
oxide and silicon at low temperatures \citep{PhysRevLett.95.210503,Krupka2006,OConnell2008}.
Yet, the high variability of the devices on native oxide points towards
an extreme sensitivity of the dielectric losses to the nanoscale variations
in the stoichiometry and thickness of the oxide.

In the rest of the paper, we will focus on the origin 
of the dephasing rates of the qubits. Indeed, the noticeable reproducibility
of the qubits enables us to analyze the different noise sources that
influence the coherence times and systematically eliminate possible
noise factors. We begin this analysis away from the optimal point,
where the flux qubit decoherence is dominated by flux noise. The power
spectrum of flux noise has a 1/f shape $S_{\varPhi}\left[f\right]=A_{\varPhi}^{2}/f$
\citep{PhysRevLett.95.257002,Bylander2011,PhysRevLett.113.123601,PhysRevB.93.104518}.
Thus, measuring the flux qubit decoherence versus $\varepsilon$ gives
us directly access to the flux noise amplitude $A_{\varPhi}$ \citep{PhysRevB.72.134519,PhysRevApplied.13.054079}.
Interestingly, we obtain almost the same flux noise amplitude $A_{\varPhi}=1.2\pm0.2\:\mu\varPhi_{0}$
for all the qubits whether on sample A, B or C including those with
constrictions or not (see \citep{sup}).

In Figure \ref{fig:rep-control-t1-t2}(c), we show the pure echo dephasing
rate $\Gamma_{2E}^{\phi}=\Gamma_{2E}-\Gamma_{1}/2$ at the optimal
point for the different qubits. At this point, the qubits are protected
against flux noise at first order. Yet, second order effects may still
impact the dephasing rates. To account for these effects, we performed
a numerical Monte Carlo simulation detailed in \citep{sup}. At the optimal
point, a simple formula is obtained: 
\[
\Gamma_{2E}^{opti}\simeq56\frac{(I_{p}A_{\varPhi}/h)^{2}}{\Delta}
\]

The results of our analysis show that second order flux noise can
only explain partially the observed dephasing at the optimal point.
Other well-known mechanisms of dephasing are related to photon noise
in the resonator \citep{PhysRevLett.95.257002,Yan2016} and charge noise \citep{PhysRevLett.113.123601}.
As shown in Figure \ref{fig:rep-control-t1-t2}(c), photon noise
has some impact on several qubits whose resonance happens to be close
to the one of the resonator. The sensitivity of flux qubits to charge-noise
is highly dependent on the ratio between the Josephson energy $E_{J}$
and the charging energy $E_{C}$. We thus calculated the maximum amplitude
of the charge modulation for each qubit (See \citep{sup}). In average,
the charge modulation is equal to 100 kHz, 5 kHz and 1 kHz for samples
A, B and C respectively. Clearly, this is more than one order of magnitude
smaller than the measured pure dephasing rate for sample B and C and
cannot explain the data. Thus, another mechanism is necessary to explain
at least qualitatively the remaining dephasing rate of these qubits.
Critical current fluctuations are for instance a possible channel
of dephasing in our system. These fluctuations are due to charges
localised in the barrier of the Josephson junctions. They also produce
a 1/f shape spectral density \citep{PhysRevLett.93.077003,Eroms2006}.
Assuming that the remaining dephasing rate of sample C is fully due
to this microscopic source of noise, we get $S_{I_{0}}[\SI{1}{\hertz}]\simeq\left(\SI{0.5}{\pico\ampere}\right)^{2}\si{\per\micro\meter\squared}$,
which seems compatible with previously reported values in the literature.

In conclusion, we have shown that flux qubits can be fabricated in
a reproducible way both in terms of gap transition energy and in terms
of decoherence rates. Reproducible relaxation times have been measured
with $T_{1}\sim15-\SI{20}{\micro\second}$ for samples fabricated
on a thermally grown 5-nm $\mathrm{SiO_{2}}$ layer. These numbers
are comparable to those observed in Ref. \citep{Yan2016} for C-shunted
flux qubits. The major advantages of our design are its large anharmonicity
($f_{12} \sim\SI{30}{\giga\hertz}$) and high persistent current ($I_{p}\sim\SI{300}{\nano\ampere}$).
This makes flux qubits ideal candidates for magnetic coupling to spins
such as NV centers \citep{PhysRevLett.105.210501,PhysRevA.92.052335}
or other impurities in silicon \citep{Albertinale2021}. In all the
samples, the amplitude of flux noise was low and reproducible $A_{\phi}=1.2\pm0.2\:\mu\varPhi_{0}/\sqrt{\mathrm{Hz}}$.
At the optimal point, long and reproducible pure dephasing times were
measured with $T_{2E}^{\phi}=15-\SI{30}{\micro\second}$. At this
level, the pure dephasing times are most likely limited by critical
current fluctuations of the small junction of the qubits. Our results
prove that flux qubits can reliably reach long coherence times and
open interesting new perspectives for both hybrid quantum circuits and scalable
quantum processing.
\begin{acknowledgments}
This research was supported by the Israeli Science Foundation under
grant numbers 426/15, 898/19 and 963/19. We acknowledge the ARC Centre
of Excellence for Quantum Computation and Communication Technology
(CE170100012). M. Stern wishes to thank fruitful discussions with
I. Bar Joseph, Y. Kubo and G. Catelani. 
\end{acknowledgments}
\begin{widetext}

\global\long\def\br#1{\left\langle #1\right|}%

\global\long\def\kt#1{\left|#1\right\rangle }%

\global\long\def\brkt#1{\left\langle #1\right\rangle }%

\global\long\def\units#1{\text{#1}}%
\global\long\def\micron{\si{\micro\meter}}%
\global\long\def\microsec{\si{\micro\second}}%
\global\long\def\identity{\mathbb{1}}%

\part*{Supplementary Materials}
\section{Experimental Setup}

Experiments are performed at a temperature of $\SI{14}{\milli\kelvin}$
in a Cryoconcept dilution refrigerator, model Hexadry 200 with low
mechanical vibrations. Supplementary Figure 1 shows a detailed schematic
of the experimental setup. The samples are glued on a microwave printed
circuit board made out of TMM10 ceramics, then enclosed in a copper
box with low mode volume which is itself embedded into a superconducting
coil that is used to provide magnetic flux biases to the qubits. To
reduce low frequency magnetic noise, the coil is surrounded by a superconducting
enclosure (Copper plated by SnPb 60/40 $\SI{15}{\micro\meter}$) and
magnetically shielded with a high permeability metal box (CryoPhy
from Meca Magnetic). The apertures of the box are tightly closed using
Eccosorb AN-72, in order to protect the sample from electromagnetic
radiation that could generate quasiparticles.\bigskip{}

The coil is powered by a BILT BE-2102 voltage source filtered by a
custom designed ultra-stable voltage to current converter. The microwaves
are generated by Keysight PSG E8257D analog microwave synthesizers.
The pulses are modulated at an intermediate frequency of 10-200 MHz
by a Quantum Machines OPX system connected to MITEQ IRM0618/IRM0408
mixers. Voltage controlled attenuators (Pulsar AAR-29-479) are used
to adjust the pulse amplitude over a wide range (0.5-64 dB). The input
line is attenuated at 4K stage (XMA -20 dB) and at the mixing chamber
stage (XMA -40 dB) to minimize thermal noise and filtered with an
homemade impedance-matched copper powder filter (-10 dB @ 10 GHz).
In addition, the pulses are shaped with smooth rise and fall ($\sim20$
ns) in order to reduce the population of microwave photons in the
resonator during coherent state evolution of the qubit.\bigskip{}

Qubit state measurement is done using dispersive readout, by measuring
the transmission of microwave pulses through the resonator, using
a custom built setup. The readout output line is filtered by two shielded
double circulators (LNF-CICIC8\_12A) and a $8-\SI{12}{\giga\hertz}$
band pass filter from Micro-Tronics, model BPC50406. The readout output
signal is amplified using a low-noise cryogenic HEMT amplifier (LNF-LNC1\_12A)
and a room temperature amplifier (LNF-LNR1\_15A). After demodulation,
the quadratures of the readout output pulse are sampled and averaged
using the IQ inputs of the OPX system. At this point, we perform a
principal axis transformation on the data points by diagonalizing
their covariance matrix. Using this transformation, we extract the
largest principal component of the measured $(I,Q)$ points and obtain
the state of the qubit.

\clearpage{}

\begin{figure}[H]
\begin{centering}
\includegraphics[width=0.85\linewidth]{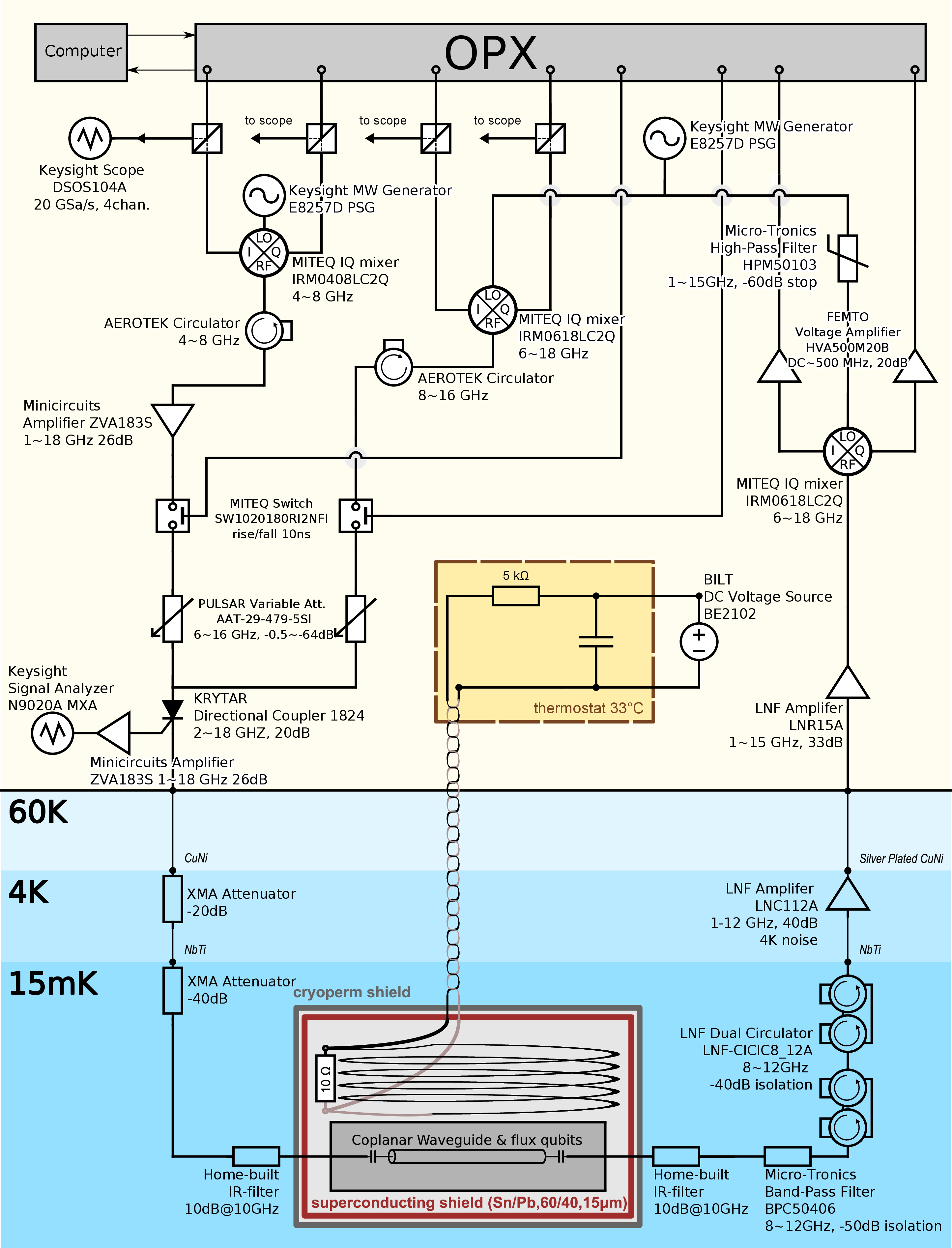}
\par\end{centering}
\caption{Experimental Setup.}
\end{figure}
\clearpage{}

\section{Flux Qubit Model}

\global\long\def\cgeom{\mathbf{C_{geom}}}%
\global\long\def\cjun{\mathbf{C_{J}}}%
\global\long\def\cedge{\mathbf{C_{edge}}}%
\global\long\def\ccoarse{\mathbf{C_{coarse}}}%
\global\long\def\cmat{\mathbf{C}}%

The flux qubit consists of a superconducting loop intersected by four
Josephson junctions among which one is smaller than others by a factor
$\alpha$. \ref{fig:circuitdiag} shows a schematic drawing of a flux
qubit. Each Josephson junction is characterized by its Josephson energy
$E_{J}$ and its bare capacitance $C_{J}$. The junctions divide the
loop into four superconducting islands. The island $I_{1}$ is galvanically
connected to the coplanar waveguide resonator. Each island is capacitively
coupled to its surrounding by geometric capacitances denoted as $C_{ij}$
where $\left(i,j\right)\in\left(0,1,..,4\right)$, the index 0 representing
the ground.
\begin{figure}[H]
\begin{centering}
\includegraphics[height=0.45\textwidth]{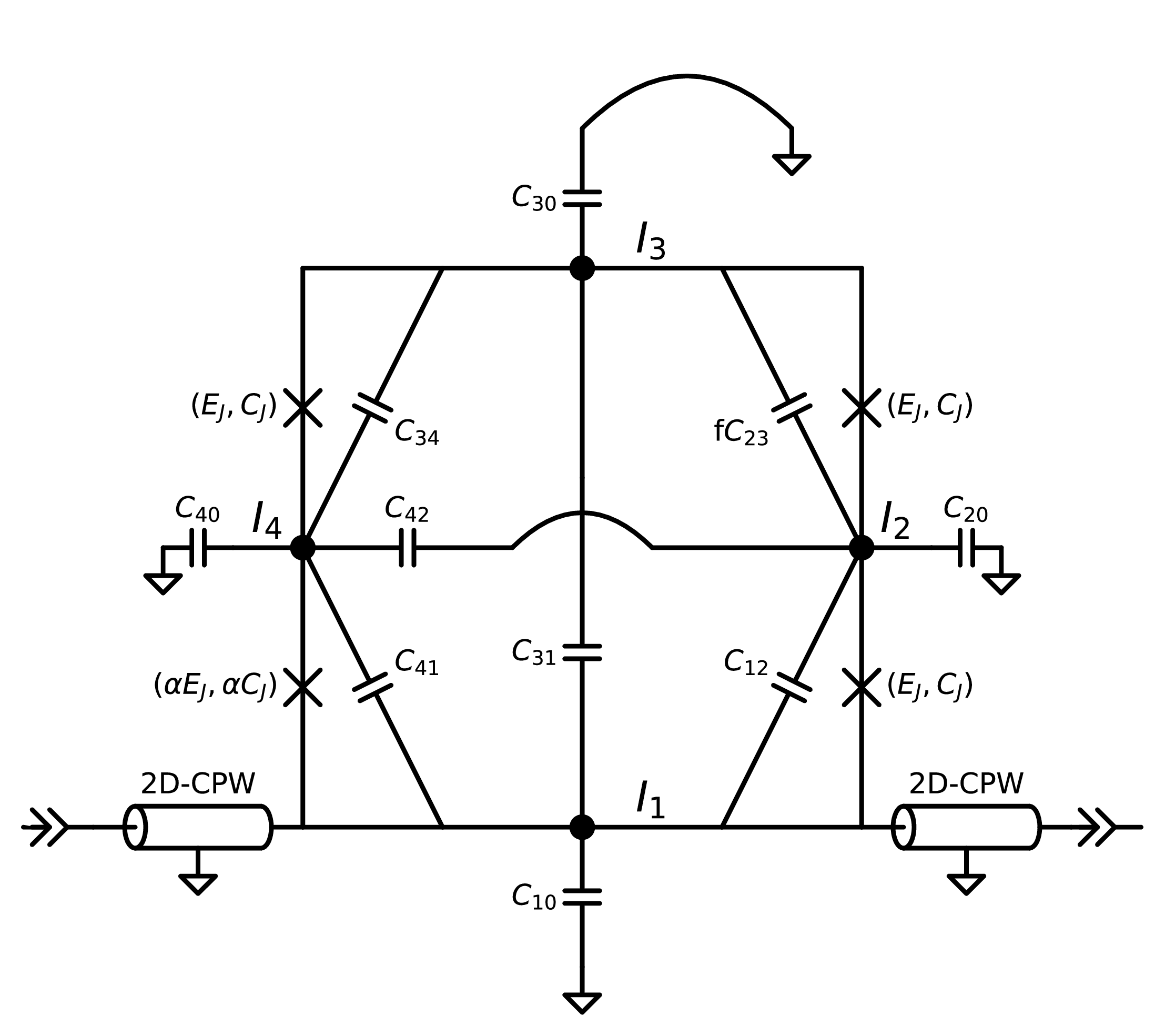}
\par\end{centering}
\caption{\textbf{Equivalent circuit diagram of a flux qubit.} The Josephson
junctions are defined by their Josephson energy $E_{J}$ and their
bare capacitance $C_{J}$. The island $I_{1}$ is galvanically connected
to the coplanar waveguide resonator. Each island is capacitively coupled
to its surrounding by geometric capacitances denoted as $C_{ij}$
where $\left(i,j\right)\in\left(0,1,..,4\right)$, the index 0 representing
the ground. \label{fig:circuitdiag}}
\end{figure}

\subsection{Potential Energy}

The potential energy of the circuit shown in \ref{fig:circuitdiag}
corresponds to the inductive energy of the junctions and can be written
as
\begin{equation}
U=-\sum_{j=1}^{3}E_{J}\cos\varphi_{j,j+1}-\alpha E_{J}\cos\varphi_{41}\label{eq:ham-pot}
\end{equation}

where $\varphi_{j,k}$ denotes the phase difference $\varphi_{k}-\varphi_{j}$
between islands $j$ and $k$. 

Faraday law implies that 
\begin{equation}
\varphi_{41}=2\pi\frac{\Phi}{\Phi_{0}}-\sum_{j=1}^{3}\varphi_{j,j+1}
\end{equation}
where $\Phi$ is the flux threading the qubit loop and $\Phi_{0}=h/2e$. 

When $\Phi=\Phi_{0}/2$, the potential energy has two degenerated
minima. The positions of these minima are given by solving the partial
differential equations $\partial_{\varphi_{i}}U=0$. The two solutions
verify the simple equation $\sin\varphi^{*}=\alpha\sin3\varphi^{*}$
and correspond to two opposite persistent currents given by 
\begin{equation}
I_{p}=\pm I_{0}\sqrt{\frac{3}{4}-\frac{1}{4\alpha}}
\end{equation}

where $I_{0}$ is the critical current of the Josephson junctions.

\subsection{Kinetic Energy}

The kinetic energy $K$ of the system is the sum of the capacitive
energies of the circuit 

\begin{equation}
K=\frac{1}{2}\sum_{i\neq j}C_{ij}\left(V_{j}-V_{i}\right)^{2}+\frac{1}{2}C_{J}\left(\left(V_{1}-V_{2}\right)^{2}+\left(V_{2}-V_{3}\right)^{2}+\left(V_{3}-V_{4}\right)^{2}+\alpha\left(V_{4}-V_{1}\right)^{2}\right)
\end{equation}

It is a quadratic form of the island voltages $V_{i}$ and can thus
be written as 

\begin{equation}
K=\frac{1}{2}\mathbf{V}^{T}\cmat\mathbf{V}
\end{equation}

where $\mathbf{V^{T}}=\left(\begin{array}{cccc}
V_{1} & ,V_{2} & ,V_{3}, & V_{4}\end{array}\right)$ and $\cmat$ is a $4\times4$ matrix which we will refer in the following
as the capacitance matrix. The matrix $\cmat$ can be written as the
sum of the Josephson capacitance matrix $\cjun$ and the geometric
capacitance matrix $\cgeom$: 

\begin{equation}
\cmat=\cjun+\cgeom
\end{equation}

where
\begin{equation}
\cjun=C_{J}\left(\begin{array}{cccc}
1+\alpha & -1 & 0 & -\alpha\\
-1 & 2 & -1 & 0\\
0 & -1 & 2 & -1\\
-\alpha & 0 & -1 & 1+\alpha
\end{array}\right)
\end{equation}

and
\begin{equation}
\cgeom=\left(\begin{array}{cccc}
C_{10}+\sum_{j\neq1}C_{1j} & -C_{12} & -C_{13} & -C_{14}\\
-C_{21} & C_{20}+\sum_{j\neq2}C_{2j} & -C_{23} & -C_{24}\\
-C_{31} & -C_{32} & C_{30}+\sum_{j\neq3}C_{3j} & -C_{34}\\
-C_{41} & -C_{42} & -C_{43} & C_{40}+\sum_{j\neq4}C_{4j}
\end{array}\right)
\end{equation}

We determined the capacitance matrix $\cgeom$ using an electrostatic
simulator (COMSOL) and according to the prescriptions detailled herein
below. 

\clearpage{}

\subsection{Numerical Estimation of the Geometrical Capacitance}

Numerical Estimation of the geometrical capacitance using finite element
solvers is difficult due to the different length scales involved.
The qubits have typically micron size dimensions while the oxide thickness
is rather of the order of 1 nm. As a consequence, a fine meshing is
difficult to establish. In this section, we will present an approach
which provides satisfactory results. 

\subsubsection{Coarse estimation}

We first performed a coarse simulation using the electrostatic module
of COMSOL. To perform this simulation, we assumed Neumann boundary
conditions (zero charge) on a box of $30\,\micron$ surrounding the
qubit (see \ref{fig:comsol-box-mesh}). The oxide of the Josephson
junctions was replaced by a hollow box of thickness $l=\SI{20}{nm}$.
We defined a minimum meshing size of $\SI{4}{nm}$. For these mesh
parameters, the far field components are accurately calculated. Isolated
islands not participating in the flux qubit loop were set to charge
conservation $Q=0$ terminal settings. 

\begin{figure}[h]
\begin{centering}
\includegraphics[width=0.7\linewidth]{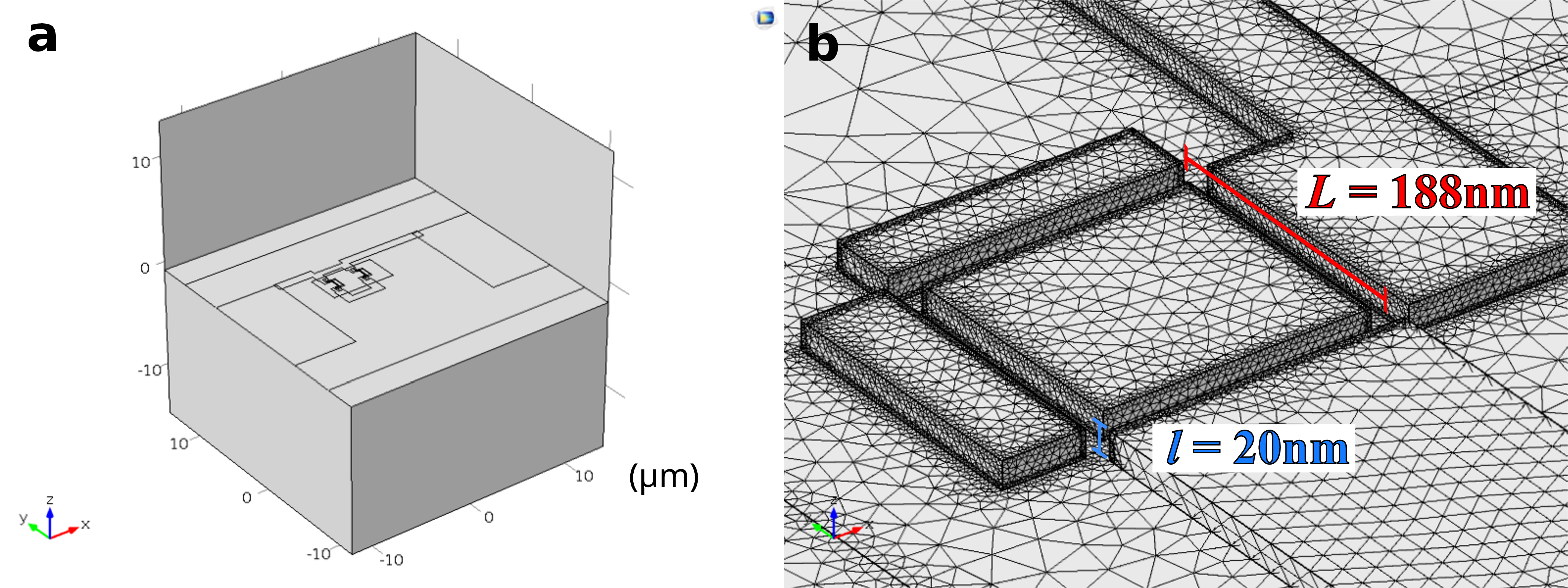}
\par\end{centering}
\caption{\textbf{Coarse Simulation using the electrostatic module of COMSOL.}
\textbf{a}, The precise design of the qubit is imported to the simulator
and put into a cubic box of $30\,\protect\micron$ edge, where zero
charge boundary condition is imposed. \textbf{b}, Close-up view of
the meshing around one of the junction. The junction is modelized
by conducting planes separated by 20 nm distance in order to keep
a minimal meshing size of 4 nm. \label{fig:comsol-box-mesh}}
\end{figure}
 We applied sequentially a voltage on each island in order to construct
the capacitance matrix $\ccoarse$. For instance, the coarse capacitance
matrix of qubit B4 is
\[
\ccoarse=\left(\begin{array}{cccc}
3.752 & -0.181 & -0.524 & -0.137\\
-0.181 & 0.350 & -0.148 & -0.002\\
-0.524 & -0.148 & 1.044 & -0.140\\
-0.137 & -0.002 & -0.140 & 0.300
\end{array}\right)\qquad\mathrm{fF}
\]

\subsubsection{Estimating the capacitance between edges}

In order to obtain more precise results, the capacitance between adjacent
edges needs to be corrected. In \ref{fig:typical-dolan-edge} , we
represent a close-up view of a typical Josephson junction obtained
by Dolan technique, where we show the four edge capacitances we need
to consider. The two capacitances $C_{\text{edge}}^{\text{Si}}$ are
dominant due to the high permittivity constant of Si and thus $C_{\text{edge}}^{\text{air}}$
can be neglected in a first approximation.

\begin{figure}[h]
\begin{centering}
\includegraphics[width=0.8\linewidth]{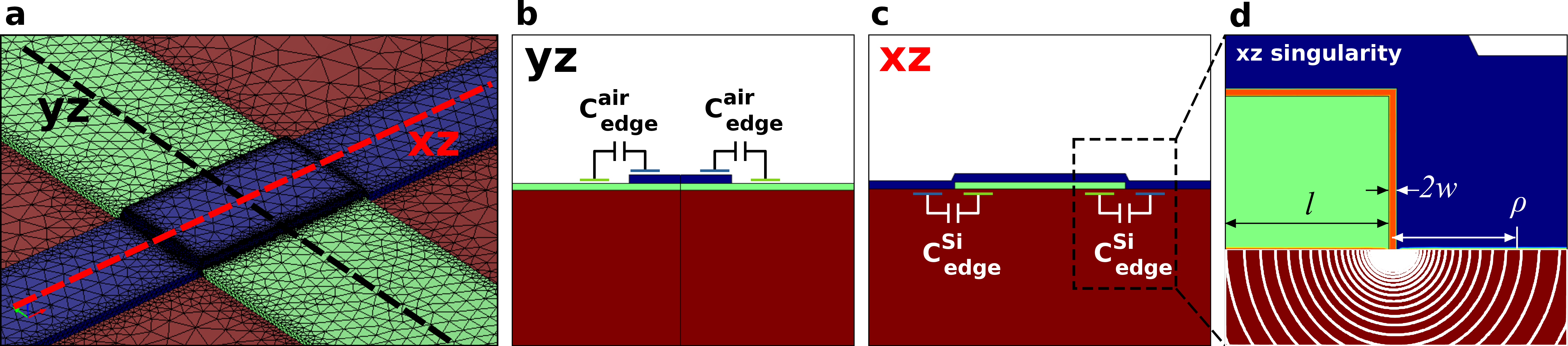}
\par\end{centering}
\caption{\textbf{Edge capacitances.} \textbf{a}, 3D representation of a Josephson
junction obtained by double angle evaporation. \textbf{b}, View cut
of the junction in the yz plane showing the $C_{edge}^{air}$ edge
capacitances. \textbf{c}, View cut of the junction in the xz plane
showing the $C_{edge}^{Si}$ edge capacitances. \textbf{d}, Close
up view of the edge showing the electric field lines giving rise to
the edge capacitances. We show the edge length $l$, the distance
to the singularity $\rho$, and the oxide thickness $2w$. \label{fig:typical-dolan-edge}}
\end{figure}

The capacitance between adjacent edges of length $l=20$ nm and width
$L$ separated by an oxide layer in the region$\left|\rho\right|<w$
(See \ref{fig:typical-dolan-edge}d) can be calculated analytically.
By using Gauss theorem, we have

\begin{equation}
L\int_{w}^{l}\frac{V}{\pi\rho}d\rho=\frac{Q}{\epsilon_{0}\epsilon_{r}}
\end{equation}
where $V$ is the voltage potential in the silicon substrate at a
distance $\rho$ from the junction singularity, $\epsilon_{0}\epsilon_{r}$
is the dielectric permittivity of silicon and $Q$ the charge accumulated
on the surface of the metallic island. Thus, the capacitance is given
by 

\begin{equation}
C_{edge}^{Si}=L\frac{\epsilon_{0}\epsilon_{r}}{\pi}\ln\frac{l}{w}
\end{equation}

For instance, the edge capacitance matrix of qubit B4 is

\[
\cedge=\left(\begin{array}{cccc}
0.069 & -0.042 & 0 & -0.027\\
-0.042 & 0.084 & -0.042 & 0\\
0 & -0.042 & 0.084 & -0.042\\
-0.027 & 0 & -0.042 & 0.069
\end{array}\right)\qquad\mathrm{fF}
\]

\subsubsection{Numerical results\label{subsec:cgeo_calc_num}}

Following the procedure described herein above, the capacitances matrix
of qubit B4 is calculated and given here as an example:

\begin{align*}
\cjun & =\left(\begin{array}{cccc}
7.898 & -5.265 & 0 & -2.633\\
-5.265 & 10.531 & -5.265 & 0\\
0 & -5.265 & 10.531 & -5.265\\
-2.633 & 0 & -5.265 & 7.898
\end{array}\right)\qquad\mathrm{fF}\\
\mathbf{\cgeom} & =\ccoarse+\cedge=\left(\begin{array}{cccc}
3.821 & -0.223 & -0.524 & -0.164\\
-0.223 & 0.434 & -0.190 & -0.002\\
-0.524 & -0.190 & 1.128 & -0.182\\
-0.164 & -0.002 & -0.182 & 0.370
\end{array}\right)\qquad\mathrm{fF}
\end{align*}

This matrix is then inserted in the Lagrangian of the qubit as we
will see herein below. 

\subsection{Legendre Transformation and Hamiltonian}

The Lagrangian of the system is $\mathcal{L}=K-U$. The conjugate
momenta of our system are given by

\begin{equation}
n_{j}\equiv\frac{1}{\hbar}\frac{\partial\mathcal{L}}{\partial\dot{\varphi}_{j,j+1}}
\end{equation}

Since $\frac{\varPhi_{0}}{2\pi}\dot{\varphi}_{j,j+1}=V_{j+1}-V_{j}$,
it is neccessary to express the kinetic energy terms in a new basis.
Since island $I_{1}$ is galvanically connected to the central conductor
of the CPW, we can safely assume that $V_{1}=0\,\mathrm{V}$, which
simplifies considerably the transformation:

\begin{align*}
V_{1} & =0\\
V_{2} & =\cancelto{0}{V_{1}}+V_{12}\\
V_{3} & =\cancelto{0}{V_{1}}+V_{12}+V_{23}\\
V_{4} & =\cancelto{0}{V_{1}}+V_{12}+V_{23}+V_{34}
\end{align*}

where $V_{ij}=V_{j}-V_{i}$. The passage matrix $P$ between these
two bases can be thus written as
\begin{equation}
\mathbf{P}=\left(\begin{array}{ccc}
0 & 0 & 0\\
1 & 0 & 0\\
1 & 1 & 0\\
1 & 1 & 1
\end{array}\right)
\end{equation}

\global\long\def\hamil{\mathcal{H}}%

The Hamiltonian $\hamil$ is then obtained by the Legendre transformation
$\mathcal{\hamil}=\hbar\sum_{j=1}^{3}\dot{\varphi}_{j,j+1}n_{j}-\mathcal{L}$
and thus writes

\begin{equation}
\hamil=\frac{\left(2e\right)^{2}}{2}n^{T}\left(\mathbf{P}^{T}\cmat\mathbf{P}\right)^{-1}n+U\label{eq:Hc-after-Leg}
\end{equation}

This Hamiltonian can be expressed in the so-called charge basis $\kt{n_{1},n_{2},n_{3}},\forall n_{1},n_{2},n_{3}\in\mathbb{Z}^{3}$,
noting that 
\begin{equation}
\cos\varphi_{j,j+1}\kt{n_{1},n_{2},n_{3}}=\frac{1}{2}\left(\kt{n_{1}+\delta_{j1},n_{2}+\delta_{j2},n_{3}+\delta_{j3}}+\kt{n_{1}-\delta_{j1},n_{2}-\delta_{j2},n_{3}-\delta_{j3}}\right)
\end{equation}
In this basis the operator $\frac{\left(2e\right)^{2}}{2}n^{T}\left(\mathbf{P}^{T}\cmat\mathbf{P}\right)^{-1}n$
is diagonal while the operator $U$ is sparse. The precision of the
eigenvalues and eigenstates depends on the truncation of the $n_{j}$
bases. With $n_{k}=-10...10$, we would need $21^{3}$ coefficients
just to describe the wavefunction and another $\left(21^{3}\right)^{2}$
to describe the Hamiltonian matrix. Thanks to the the sparsity of
the Hamiltonian operator, the number of nonzero entries in this matrix
is only $21^{3}\times\left(1+4\times2\right)$. This resolution in
charge space is computationally feasible both to store and diagonalize
matrices efficiently. For reaching the necessary precision to resolve
charge modulation, we used $n_{k}=-14...14$ and verified carefully
the numerical convergence of the calculation.

\subsection{Pseudo-Hamiltonian}

Following the full diagonalization of the Hamiltonian, we obtain the
spectrum of the flux qubit by subtracting the energy of the first
excited state $\kt 1$ from the energy of the ground state $\kt 0$.
It can be shown that close to $\Phi=\Phi_{0}/2$, the system behaves
as a two level system and the spectrum can be fully described by two
parameters:
\begin{itemize}
\item The value of the persistent current $I_{p}$, already discussed previously.
\item The so-called flux qubit gap, denoted as $\Delta$, which corresponds
to the tunneling term between the two potential minima. 
\end{itemize}
The value of the gap can be directly measured by the transition energy
at half a flux quantum $\Phi=\Phi_{0}/2$. This point is known as
the \emph{optimal point }of the flux qubit due to its immunity at
first order in flux noise, as will be explained in later sections.
In the vicinity of the optimal point, the Hamiltonian of the system
can be written using perturbation theory as

\begin{equation}
\begin{array}{c}
\hamil=\hamil_{0}-\alpha E_{J}\partial_{\Phi}\left(\text{cos}(2\pi\frac{\Phi}{\Phi_{0}}-\sum_{j=1}^{3}\varphi_{j,j+1})\right)_{\Phi=\Phi_{0}/2}\cdot\left(\Phi-\frac{\Phi_{0}}{2}\right)\\
=\hamil_{0}+\frac{1}{\varphi_{0}}\left[\underbrace{\alpha E_{J}\text{sin}\left(\varphi_{41}\right)}_{\hat{I}\cdot\varphi_{0}}\left(\Phi-\frac{\Phi_{0}}{2}\right)\right]=\hamil_{0}+\hat{I}\cdot\left(\Phi-\frac{\Phi_{0}}{2}\right)
\end{array}\label{eq:Pertobution theory}
\end{equation}

When the current operator is projected on the eigenstates $\left|0\right\rangle ,\left|1\right\rangle $
of $\hamil_{0}$ we get

\begin{equation}
\begin{array}{ccc}
\left\langle 0\right|\hat{I}\left|0\right\rangle =0 & , & \left\langle 0\right|\hat{I}\left|1\right\rangle =I_{p}\\
\left\langle 1\right|\hat{I}\left|0\right\rangle =I_{p} & , & \left\langle 1\right|\hat{I}\left|1\right\rangle =0
\end{array}\label{eq proof of the epsilon part}
\end{equation}

Therefore, the Hamiltonian of the system can be written in this basis
as

\begin{equation}
\mathcal{H}_{\text{eff}}=\frac{h}{2}\left[\Delta\sigma_{z}+\varepsilon\sigma_{x}\right]\label{eq: Heff}
\end{equation}

where $\varepsilon=\frac{2I_{p}}{h}\left(\Phi-\frac{\Phi_{0}}{2}\right)$.

The frequency of the qubit is thus given by 
\begin{equation}
\frac{\omega_{01}}{2\pi}=\sqrt{\Delta^{2}+\varepsilon^{2}}
\end{equation}

\subsection{Doublet at optimal point}

Some of the measured qubits exhibit a doublet line shape at optimal
point. This lineshape is manifested as a beating of the Ramsey oscillations
as shown in \ref{fig:Ramsey-Beating}. For qubit $B5$, the frequency
of this beating is $340\:\mathrm{kHz}$, almost two orders of magnitude
larger than the charge modulation $\delta\Delta^{n_{g}}=4.3\:\mathrm{kHz}$
and thus cannot be attributed to slow fluctuations of the electron
number parity on one of the qubit's islands \citep{PhysRevLett.113.123601,PhysRevB.91.195434}.
An alternative explanation for the origin of this doublet is related
to trapping and un-trapping of a single quasiparticle in the $\alpha$
junction.

By simple arguments, we can give a rough estimate for this effect.
The area of the $\alpha$ junction of qubit $B5$ is $A_{\alpha}=0.0257\:\mathrm{\mu m^{2}}$while
the Fermi wavelength of electrons in Aluminium is $\lambda_{F}=0.36\:\mathrm{nm}.$
Thus, the number of channels in such a junction is large and can
be estimated as $A_{\alpha}/\lambda_{F}^{2}\sim2\times10^{5}$. Assuming
that all channels have the same transmission $\tau$, we can estimate
the change of the Josephson energy of the $\alpha$ junction to be
around $1\:\mathrm{MHz}.$ We then calculate numerically the variation
$\delta\Delta^{\text{trapping}}$ of the qubit gap and obtain 300
kHz , which is close to the observed value of the doublet. We thus
come to the conclusion that these doublets are most likely due to
the trapping and un-trapping of a quasiparticle in the $\alpha$ junction.

\begin{figure}[H]
\begin{centering}
\includegraphics[width=0.5\linewidth]{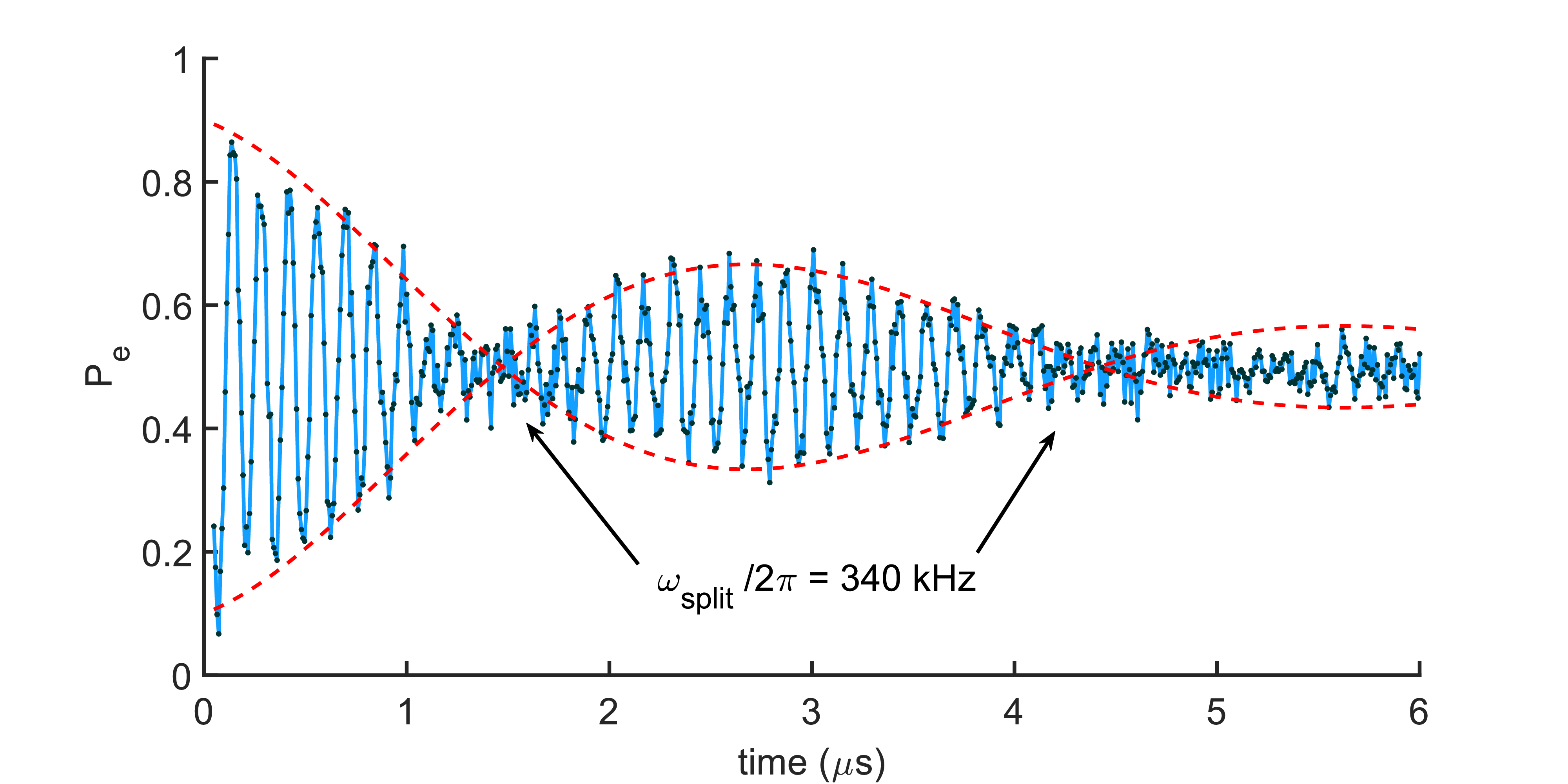}
\par\end{centering}
\caption{\textbf{Ramsey measurement of qubit B5 showing beating. \label{fig:Ramsey-Beating}}}

\end{figure}

\clearpage{}

\section{Estimating dephasing due to flux noise}

\subsection{Pure Dephasing of a flux qubit}

In an ideal system, the decoherence rate $\Gamma_{2}$ is limited
by the energy relaxation rate of the qubit and is given by $\Gamma_{2}=\Gamma_{1}/2$
. In practice, the decoherence rate of a qubit may be much larger
than this theoretical limit. There are several known sources of dephasing
which are responsible for this. Among them, flux noise, charge noise
and photon noise in the resonator. The pure dephasing rate of the
flux qubit can be estimated by the so-called Ramsey sequence, where
two identical $\pi/2$ pulses are played consecutively with a time
delay $t$. It is possible to dynamically decouple the noise responsible
for this dephasing by playing a more complex set of pulses. The most
popular technique to achieve this is called Hahn Echo technique and
consists of playing a $\pi$-pulse in between the two $\pi/2$ pulses.
This $\pi$ pulse inverses the time evolution and therefore cancels
the contribution to dephasing of low frequency noise.

In the Ramsey sequence, the first $\pi/2$-pulse raises the qubit
initially in its ground state into a coherent superposition of $\left|\Psi(0)\right\rangle =\left(\left|0\right\rangle +\left|1\right\rangle \right)/\sqrt{2}$
. During time $t$, the qubit performs a free evolution and accumulates
phase $\varphi(t)$ and becomes $\left|\Psi(t)\right\rangle =\left(\left|0\right\rangle +e^{i\varphi(t)}\left|1\right\rangle \right)/\sqrt{2}$.
The phase $\varphi(t)$ consists of two parts $\varphi(t)=\omega_{01}t+\delta(t)$,
where $\delta(t)$ is the phase due to the small fluctuations $\delta\lambda(t)$
which slightly modify the qubit Hamiltonian. At first order, $\delta(t)$
is given by $\delta(t)=\frac{\partial\omega_{01}}{\partial\lambda}\int_{0}^{t}\delta\lambda(t')dt'$.
The decoherence rate of the system corresponds to the decay of the
expectation value $\left\langle \sigma_{x}(t)\right\rangle $ and
is given by  
\[
\left\langle \sigma_{x}(t)\right\rangle =1/2\,\left\langle e^{i\varphi(t)}+e^{-i\varphi(t)}\right\rangle 
\]

When repeating the measurements, the value of $\left\langle \sigma_{x}(t)\right\rangle $
is changed due to the varying environmental noise $\delta(t)$. Therefore,
one should average the value of $e^{\pm i\delta(t)}$ in order to
determine the influence of this noise. If the fluctuations $\delta\lambda(t')$
are small enough, they can be considered as a random variable with
Gaussian distribution \citep{PhysRevB.72.134519}. Thus,

\[
f_{R}(t)=\left\langle e^{\pm i\delta(t)}\right\rangle \approx\left\langle 1\pm\cancel{i\delta}-\delta^{2}/2\right\rangle =e^{-1/2\left\langle \delta^{2}\right\rangle }
\]
 The expectation value of $\left\langle \sigma_{x}(t)\right\rangle $
will therefore decay according to
\begin{align}
f_{R}(t) & =e^{-1/2\left(\frac{\partial\omega_{01}}{\partial\lambda}\right)^{2}\left\langle \left(\int_{0}^{t}\delta\lambda(t')dt\right)^{2}\right\rangle }\\
 & \text{= exp}\left(-\frac{t^{2}}{2}\left(\frac{\partial\omega_{01}}{\partial\lambda}\right)^{2}\stackrel[-\infty]{\infty}{\int}d\omega\,S_{\lambda}(\omega)\,\text{sinc}^{2}(\frac{\omega t}{2})\right)
\end{align}

In a Hahn echo sequence, the first $\pi/2$-pulse puts the state of
the qubit in a coherent superposition state $\left|\Psi(0)\right\rangle =\left(\left|0\right\rangle +\left|1\right\rangle \right)/\sqrt{2}$
. During the time $t_{1}$, the qubit performs a free evolution and
accumulates phase $\varphi_{1}(t_{1})=\omega_{01}t_{1}+\delta_{1}(t_{1})$.
The $\pi$ - pulse flips the time evolution of the qubit such that
during the time $t_{2}$ it acquires an opposite phase $\varphi_{2}(t_{2})=-\omega_{01}t_{2}-\delta_{2}(t_{2})$.
The phase accumulated by $\omega_{01}t_{1}$ and $\omega_{01}t_{2}$
is canceled when $t_{1}=t_{2}=t/2$ and the decoherence rate of the
qubit - corresponding to the decay $f_{E}(t)=\left\langle \sigma_{x}(t)\right\rangle $
- is given by

\[
f_{E}(t)=\left\langle e^{\pm i(\delta_{1}-\delta_{2})}\right\rangle \approx\text{exp}\left(-1/2\left\langle \delta_{1}^{2}+\delta_{2}^{2}-\delta_{1}\delta_{2}-\delta_{2}\delta_{1}\right\rangle \right)
\]

The expectation value of $\left\langle \sigma_{x}(t)\right\rangle $
will therefore decay according to

\begin{equation}
f_{E}(t)=\text{exp}\left(-\frac{t^{2}}{2}\left(\frac{\partial\omega_{01}}{\partial\lambda}\right)^{2}\stackrel[-\infty]{\infty}{\int}d\omega\,S_{\lambda}(\omega)\,\text{sin}^{2}(\frac{\omega t}{4})\,\text{sinc}^{2}(\frac{\omega t}{4})\right)\label{eq:echo_decay_fac}
\end{equation}

\subsection{Dephasing away from the optimal point}

Away from the optimal point , the high magnetic moment of the circuit
($\sim500\,\mathrm{GHz/G})$ make its frequency very sensitive to
flux 
\[
\partial_{\Phi}\omega_{01}=\frac{\partial\varepsilon}{\partial\varPhi}.\frac{\partial\omega_{01}}{\partial\varepsilon}=\left(\frac{2I_{p}}{\hslash}\right)^{2}\frac{\left(\Phi-\Phi_{0}/2\right)}{\omega_{01}}
\]

The power spectrum of flux noise has a 1/f shape $S_{\varPhi}\left[\omega\right]=A_{\varPhi}^{2}/\omega$.
For the Echo sequence, one can calculate exactly the integral given
in \ref{eq:echo_decay_fac} without any additionnal assumption or approximation
and one obtains \citep{PhysRevB.72.134519}

\begin{equation}
\Gamma_{2E}^{\varphi}\left(\Phi\right)=\left(\frac{2I_{p}}{\hbar}\right)^{2}\frac{\Phi-\Phi_{0}/2}{\omega_{01}\left(\Phi\right)}A_{\Phi}\sqrt{\ln2}\label{eq:echo-1st-order}
\end{equation}

This formula is used in the following to extract the amplitude of
the flux noise as shown in \ref{fig:fn-amp}.

\begin{figure}[H]
\begin{centering}
\includegraphics[width=0.9\linewidth]{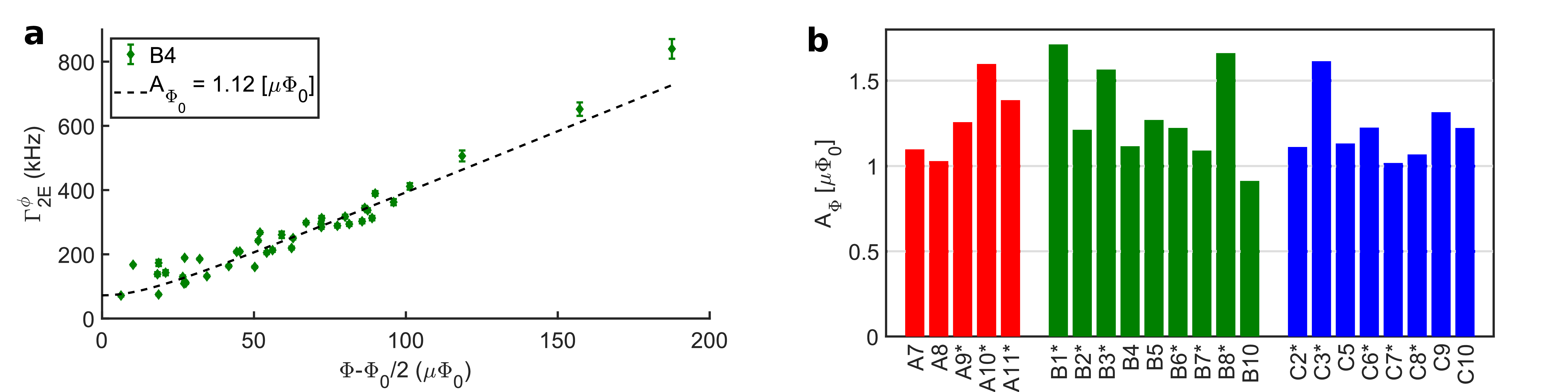}
\par\end{centering}
\caption{\textbf{Flux noise amplitude.} \textbf{a}, Echo pure dephasing rates
of qubit B4 as a function of $\varPhi-\varPhi_{0}/2$. Fitting the
measured data with \ref{eq:echo-1st-order} yields $A_{\Phi}=1.12\:\mu\Phi_{0}$
. \textbf{b}, Extracted amplitude of flux noise for different qubits
in red for sample A, green for sample B and blue for sample C. We
obtain almost the same flux noise amplitude $A_{\varPhi}=1.2\pm0.2\:\mu\varPhi_{0}$
for all the qubits whether on sample A , B or C and including those
with constrictions. \label{fig:fn-amp}}
\end{figure}

\subsection{Dephasing at the optimal point}

At the optimal point however, $\partial_{\Phi}\omega_{01}=0$ and
therefore the qubit is immune to flux fluctuations to first order.
Yet, $\partial_{\Phi}^{2}\omega_{01}=\left(\frac{2I_{p}}{\hbar}\right)^{2}/\left(2\pi\Delta\right)\neq0$
and therefore second order flux noise should be taken into account.
Unlike first order, deriving an analytical expression for 2nd order
flux noise is not straight-forward. In this work, we performed numerical
Monte Carlo simulations in Python \citep{PhysRevApplied.17.024057}.
The source code of this simulation can be found on Github \citep{Chang_pink_flux_noise_analysis_2022}. 

\subsubsection{Generation of flux noise trajectories}

\global\long\def\nfft{n_{\text{FFT}}}%

Microscopically, the flux noise is the sum of many independent uncorrelated
sources, most likely spins on the surface of the loops \citep{PhysRevApplied.13.054079}.
Thus, it should be well described as a Gaussian variable. To simulate
a flux noise trajectory in time, we first generate a series of $2\,\nfft$
normally distributed real and imaginary random numbers $a_{i}+jb_{i}$
that will be used as Fourier components of the signal. These Fourier
components are multiplied by an amplitude $A_{\Phi}\sqrt{\frac{\nfft dt}{i}}$
where $dt$ is the time step unit of the simulation. Then, we apply
an inverse fast Fourier transform in order to obtain a flux noise
trajectory with power spectrum of $A_{\Phi}^{2}/f$. In the code,
the class \code{NoiseGen1OverF} is a generator of pink noise. Attributes
includes the time step unit $dt$, and the total number $n_{\text{FFT}}$
of samples to generate. The method \code{generate} is called to generate
a single trajectory $\overrightarrow{\delta\Phi}=\left(\delta\Phi\left(t_{1}\right),\ldots,\delta\Phi\left(t_{\nfft}\right)\right)$
around zero flux.

\subsubsection{Ensemble averaging over noise trajectories}

The next step of our simulation consists of ensemble averaging of
a complex function over different trajectories. For Ramsey sequence,
this complex function is 
\[
f_{R}\left(t\right)=\exp\left(i\int_{0}^{t}\omega_{01}\left(\Phi+\delta\Phi\left(u\right)\right)-\left\langle \omega_{01}\right\rangle du\right)
\]
, where $\Phi$ is the flux threading the loop of the qubit. For Hahn-Echo
sequence, the complex function is 
\[
f_{E}\left(t\right)=\exp\left(i\int_{0}^{t/2}\omega_{01}\left(\Phi+\delta\Phi\left(u\right)\right)du-i\int_{t/2}^{t}\omega_{01}\left(\Phi+\delta\Phi\left(u\right)\right)du\right)
\]

In order to reduce the function call overheads, the function \code{qb\_plot\_t2s\_at}
performs the ensemble averaging by sampling the $n_{\text{FFT}}$-sized
signal at fixed intervals, much like in a real experiment where ensemble
repetitions occur in sequential order at a quasi-fixed period. To
further increase the smoothness of the signal, we resample the same
signal using the same period but with different time offsets.

\begin{lstlisting}[language=Python,tabsize=4]
for i in tqdm.tqdm(range(0, tobs_tstep, laziness)):#line 166
	...# for different periodic offsets
\end{lstlisting}
In order to optimize the running complexity, the integral $\int_{t_{i}}^{t_{j}}\omega_{01}\left(\Phi+\delta\Phi\left(u\right)\right)du$
is calculated  as a difference of pre-cached cumulative sums $\int_{0}^{t}\omega_{01}\left(\Phi+\delta\Phi\left(u\right)\right)du$:

\[
\int_{t}^{t_{j}}\omega_{01}\left(\Phi+\delta\Phi\left(u\right)\right)du=\int_{0}^{t_{j}}\omega_{01}\left(\Phi+\delta\Phi\left(u\right)\right)du-\int_{0}^{t_{i}}\left(\Phi+\delta\Phi\left(u\right)\right)du
\]

The pre-caching step is performed in only $O\left(n_{\text{FFT}}\right)$
complexity. To further speed up the whole algorithm, we perform the
computation described above by using \code{np.reshape} commands instead
of writing python for-loops, to exploit the faster speed of C-implemented
\code{numpy} libraries.

Here is a list of important arguments of the function \code{qb\_plot\_t2s\_at}:
\begin{enumerate}
\item \code{t\_step\_ns} corresponds to the time step unit $dt$
\item \code{t\_observation\_ns}, time interval on which the interpulse
time delay will be varied. This is the X axis of the final plot.
\item \code{t\_total\_ns}, this is the total length of the pink signal,
equal to $\nfft dt$. The inverse is the resolution $df$ in frequency
space.
\item \code{t\_cut\_off\_ns}. Its inverse is the low frequency cutoff of
the power spectrum. We assume white noise below this threshold.
\end{enumerate}
The default sample program provided under the \code{\_\_main\_\_}
statement performs the following steps. First a typical flux qubit
transition with parameters $\Delta,I_{p}$, under influence of pink
noise of amplitude $A_{\Phi_{0}}$ is defined. Sanity checks on the
calculations of the qubit's first and second derivatives are performed
(cf. equality \code{c1 == c1\_sp} and \code{c2 == c2\_sp}). Finally,
after the averaging is complete, a plot of the ensemble averaged signal
should pop up. The titles prints the decoherence times $\tau_{2E/R}$,
defined by $\left|c\left(\tau_{2E/R}\right)\right|=1/e$.

\begin{figure}[h]
\centering{}\includegraphics[width=0.9\textwidth]{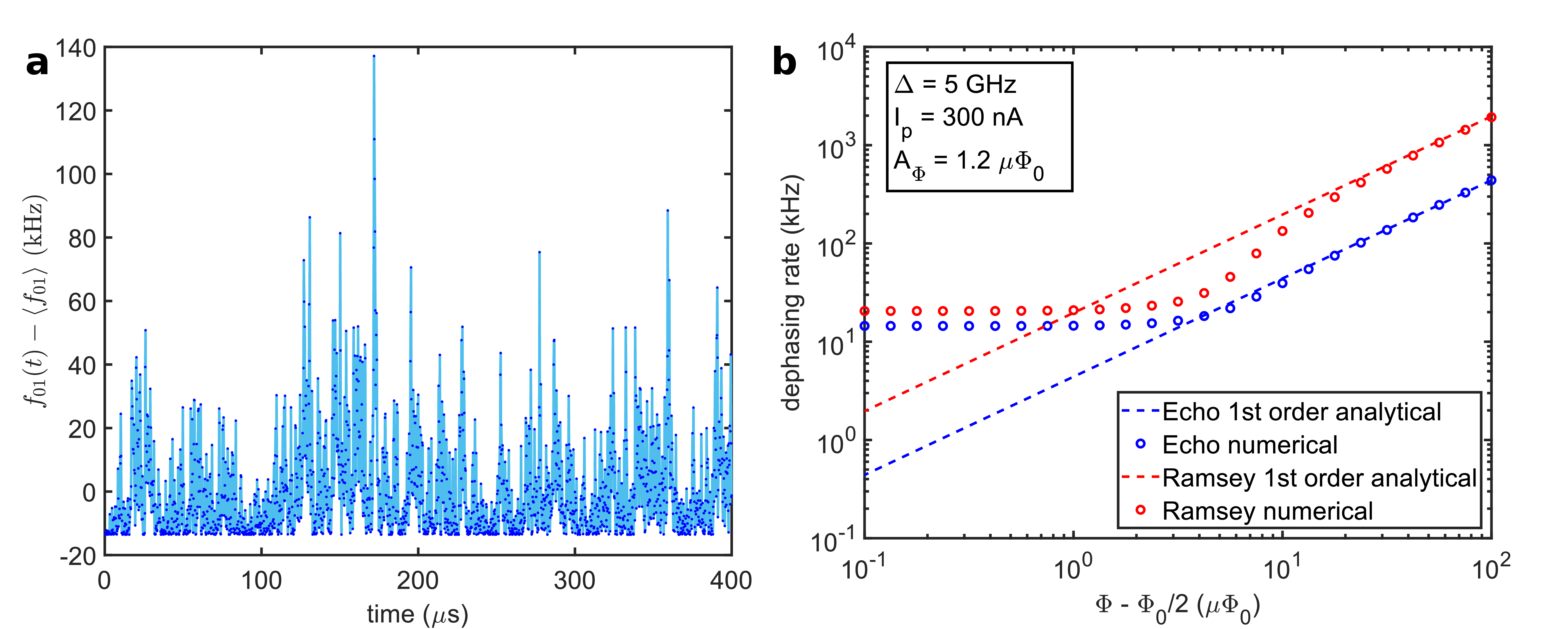}\caption{\textbf{Calculated dephasing rates of a flux qubit. a}, The frequency
of a flux qubit at optimal point under influence of $1/f$ noise over
a sub-sample of $400\ \protect\microsec$. The flux qubit parameters
were chosen to be $\Delta=5\:\mathrm{GHz},I_{p}=300\:\mathrm{nA},A_{\varPhi}=1.2\:\mu\Phi_{0}$.
The parameters of the numerical simulation are $t_{\text{step}}=200\text{ ns}$,
$t_{\text{total}}=1\text{ s}$, $t_{\text{cut-off}}=0.2\text{ s}$.
\textbf{b}, We numerically calculated the Ramsey and Echo dephasing
rates and compared the results with analytical formula for first order
flux noise. Away from the optimal point, analytical formula predict
$\Gamma_{\varphi R}\sim4.5\:\Gamma_{\varphi E}$ in agreement with
the numerical simulations.\label{fig:def-plot-fn}}
\end{figure}

\subsubsection{Empirical results for the second-order flux noise decoherence rates}

Using the tool described above, and sweeping many different flux qubit
parameters, we were able to establish the following empirical law
for any second-order transition

\begin{equation}
\Gamma_{\varphi E}^{\left(2\right)}=14.4\frac{\partial^{2}f_{01}}{\partial\Phi^{2}}A_{\Phi_{0}}^{2}
\end{equation}

where $f_{01}$ is the transition frequency.

For the particular case of the flux qubit at its optimal point, we
obtain the formula used in the main article

\begin{equation}
\Gamma_{\varphi E}^{\left(2\right)}=56\frac{\left(I_{p}A_{\Phi_{0}}/h\right)^{2}}{\Delta}
\end{equation}

\section{Qubit and resonator parameters \label{sec:Qubit-Parameters-Table}}

\begin{table}[H]
\begin{centering}
\begin{tabular}{|c|c|c|c|c|c|c|c|c|c|}
\hline 
﻿Qubit ref. & $\Delta$ (GHz) & $I_{p}$ (nA) & $g$ (MHz) & $\chi$ (MHz)  & $E_{J}$ (GHz) & $E_{J}/E_{C}$ & $\alpha$ & $\delta\Delta^{\text{geom}}$ (GHz) & $\delta\Delta^{n_{g}}$ (kHz)\tabularnewline
\hline 
A2 & 7.19 & 188 & 28 & 1.29 & 265 & 69 & 0.492 & -1.09 & 78.4\tabularnewline
A6 & 8.33 & 244 & 50 & 4.48 & 256 & 67 & 0.487 & -0.51 & 95.1\tabularnewline
A7 & 8.69 & 187 & 40 & 1.8 & 240 & 63 & 0.476 & -1.34 & 136.3\tabularnewline
A8 & 6.35 & 202 & 43 & 1.19 & 264 & 69 & 0.504 & -0.83 & 94.5\tabularnewline
A9 & 5.24 & 201 & 50 & 0.82 & 251 & 66 & 0.514 & -1.25 & 154.6\tabularnewline
A10 & 8.35 & 182 & 52 & 4.66 & 255 & 67 & 0.477 & -1.54 & 88.1\tabularnewline
A11 & 5.81 & 201 & 61 & 1.64 & 258 & 67 & 0.503 & -1.49 & 110.1\tabularnewline
\hline 
B1 & 5.73 & 289 & 94 & 1.59 & 362 & 98 & 0.489 & -1.61 & 3.7\tabularnewline
B2 & 4.48 & 302 & 95 & 1.05 & 360 & 98 & 0.504 & -1.36 & 5.0\tabularnewline
B3 & 4.01 & 310 & 92 & 0.86 & 361 & 98 & 0.511 & -1.24 & 5.4\tabularnewline
B4 & 5.18 & 298 & 79 & 0.94 & 364 & 99 & 0.497 & -1.30 & 4.0\tabularnewline
B5 & 5.84 & 284 & 77 & 1.11 & 357 & 97 & 0.490 & -1.46 & 4.3\tabularnewline
B6 & 5.08 & 297 & 71 & 0.72 & 361 & 98 & 0.499 & -1.27 & 4.5\tabularnewline
B7 & 5.01 & 292 & 64 & 0.57 & 354 & 96 & 0.500 & -1.31 & 5.4\tabularnewline
B8 & 3.88 & 316 & 63 & 0.37 & 366 & 100 & 0.512 & -1.17 & 4.8\tabularnewline
B10 & 5.69 & 288 & 40 & 0.29 & 360 & 98 & 0.490 & -1.53 & 4.0\tabularnewline
\hline 
C2 & 5.93 & 303 & 90 & 1.56 & 387 & 111 & 0.481 & -1.54 & 1.0\tabularnewline
C3 & 5.74 & 304 & 85 & 1.29 & 386 & 110 & 0.484 & -1.47 & 1.1\tabularnewline
C5 & 6.80 & 288 & 75 & 1.5 & 380 & 109 & 0.474 & -1.52 & 1.1\tabularnewline
C6 & 7.39 & 285 & 64 & 1.43 & 383 & 110 & 0.470 & -1.50 & 0.9\tabularnewline
C7 & 7.16 & 295 & 67 & 1.4 & 394 & 113 & 0.470 & -1.54 & 0.7\tabularnewline
C8 & 7.30 & 287 & 58 & 1.11 & 386 & 110 & 0.469 & -1.61 & 0.8\tabularnewline
C9 & 6.14 & 295 & 55 & 0.63 & 380 & 109 & 0.480 & -1.48 & 1.2\tabularnewline
C10 & 6.40 & 300 & 42 & 0.41 & 391 & 112 & 0.477 & -1.57 & 0.8\tabularnewline
\hline 
\end{tabular}
\par\end{centering}
\caption{\textbf{Qubit parameters. }The values of $\Delta,$$I_{p}$, $g$
and $\chi$ were extracted from fit of the data as described in the
main text. We simulated the design of each qubit with the electrostatic
simulator of COMSOL in order to obtain the geometric capacitance matrix
of the system. This matrix was corrected according to the prescriptions
described herein above. Then, we fitted the parameters $E_{J}$ and
$\alpha$ for each qubit assuming that the junctions have a capacitance
per unit area $Cc=100\:\mathrm{fF/\mu m^{2}}$. The capacitance energy
of the junctions is defined as $E_{c}=\frac{e^{2}}{2C_{J}}$. $\delta\Delta^{\text{geom}}$
is the difference between the value of the gap calculated with and
without taking into account the geometric capacitance. As a rule of
thumb, the geometric capacitance reduces the gap of the qubit by approximately
1-1.5 GHz. $\delta\Delta^{n_{g}}$is the charge modulation calculated
for the fitted parameters $E_{J}$ and $\alpha$ of each qubit.}
\end{table}

\begin{table}[H]
\begin{centering}
\begin{tabular}{|c|c|c|c|c|c|c|c|c|}
\hline 
﻿Resonator & Length ($\micron$) & $C_{C}$ (fF) & $f_{r}$ (GHz) & $Q_{tot}$ & $\kappa\:\left(\mathrm{rad.s^{-1}}\right)$ & $Q_{C}$ & $Q_{int}$ & $\bar{n}_{thermal}$\tabularnewline
\hline 
A & 7250 & \multirow{3}{*}{5} & 7.756 & 1400 & $3.5\times10^{7}$ & 5500 & 1878 & $8.88\times10^{-4}$\tabularnewline
\cline{1-2} \cline{2-2} \cline{4-9} \cline{5-9} \cline{6-9} \cline{7-9} \cline{8-9} \cline{9-9} 
B & 5730 &  & 9.805 & 2800 & $2.2\times10^{7}$ & 3500 & 14000 & $6.82\times10^{-4}$\tabularnewline
\cline{1-2} \cline{2-2} \cline{4-9} \cline{5-9} \cline{6-9} \cline{7-9} \cline{8-9} \cline{9-9} 
C & 5730 &  & 9.850 & 2200 & $2.8\times10^{7}$ & 3500 & 5923 & $6.78\times10^{-4}$\tabularnewline
\hline 
\end{tabular}
\par\end{centering}
\caption{\textbf{Resonator parameters.} Length of the resonator, coupling capacitance
$C_{C}$, bare frequency of the resonator $f_{r}$, quality factor
$Q$ of the resonator, photon loss rate $\kappa$, coupling $Q_{C}$
and internal $Q_{int}$ quality factor of the resonator given by $\frac{1}{Q}=\frac{1}{Q_{C}}+\frac{1}{Q_{int}}$,
estimated number of thermal photons in the resonator $\bar{n}_{thermal}$.}
\end{table}

\begin{table}[H]
\begin{centering}
\begin{tabular}{|c|c|c|c|c|c|}
\hline 
Qubit ref. & $\Gamma_{1}$ (kHz) & $\Gamma_{P}$ (kHz) & $\Gamma_{2E}^{\phi}$ (kHz) & $\Gamma_{2E}^{2nd}$ (kHz) & $\Gamma_{2E}^{photon}$ (kHz)\tabularnewline
\hline 
A2 & 1363 & 36 & x & x & 10\tabularnewline
A6 & 354 & 143 & x & x & 131\tabularnewline
A7 & 99 & 96 & 99 & 3 & 21\tabularnewline
A8 & 38 & 17 & 46 & 4 & 9\tabularnewline
A9 & 31 & 4 & 97 & 7 & 4\tabularnewline
A10 & 330 & 245 & 440 & 6 & 134\tabularnewline
A11 & 37 & 25 & 131 & 8 & 16\tabularnewline
\hline 
B1 & 84 & 4 & 169 & 24 & 12\tabularnewline
B2 & 78 & 1 & 151 & 17 & 5\tabularnewline
B3 & 82 & 0 & 123 & 33 & 4\tabularnewline
B4 & 52 & 1 & 67 & 12 & 4\tabularnewline
B5 & 67 & 1 & 115 & 13 & 6\tabularnewline
B6 & 43 & 1 & 51 & 15 & 2\tabularnewline
B7 & 55 & 1 & 217 & 11 & 1\tabularnewline
B8 & 41 & 1 & 113 & 40 & 1\tabularnewline
B10 & 59 & 3 & 113 & 7 & 0\tabularnewline
\hline 
C2 & 72 & 0 & 39 & 11 & 6\tabularnewline
C3 & 54 & 2 & 77 & 24 & 4\tabularnewline
C5 & 66 & 0 & 60 & 9 & 5\tabularnewline
C6 & 82 & 1 & 78 & 9 & 5\tabularnewline
C7 & 71 & 0 & 35 & 7 & 4\tabularnewline
C8 & 86 & 12 & 42 & 7 & 3\tabularnewline
C9 & 48 & 0 & 54 & 14 & 1\tabularnewline
C10 & 67 & 0 & 32 & 12 & 0\tabularnewline
\hline 
\end{tabular}
\par\end{centering}
\caption{\textbf{Decoherence rates. }Measured relaxation rate $\Gamma_{1},$calculated
Purcell decay rate $\Gamma_{P}$, measured pure echo dephasing rate
at optimal point $\Gamma_{2E}^{\phi}$, calculated second order flux
noise dephasing rate $\Gamma_{2E}^{2nd}$, calculated photon dephasing
rate $\Gamma_{2E}^{photon}\approx4\,\frac{\left(2\pi\chi\right)^{2}}{\kappa}\bar{n}_{thermal}\left(\bar{n}_{thermal}+1\right)$.}
\end{table}
\clearpage{}

\section{Qubit fabrication and Room Temperature resistance measurements}

The samples were fabricated on a $\SI{300}{\micro\meter}$ thick wafer
of intrinsic silicon (resistivity $>10000\,\mathrm{\Omega}.\mathrm{cm}$)
for sample A and on a thermally grown 5-nm width $\mathrm{SiO_{2}}$
layer for sample B and C. The oxide layer was grown at $\SI{800}{\celsius}$
for 20 minutes. This was immediately followed by a 60 minute anneal
in nitrogen at the same temperature to reduce the fixed oxide charge.
A $\SI{400}{\celsius}$ anneal in forming gas (Ar/H) concluded the
process which was designed to passivate dangling bonds at the Si-$\mathrm{SiO_{2}}$
interface. Capacitance measurements on test devices yield an interface
state density in the low $10^{10}\si{\per\electronvolt\per\centi\meter\squared}$.
The fixed oxide charge is estimated to be in a similar range. 

The silicon wafer was dipped into Piranha acid ($\mathrm{H_{2}SO_{4}:H_{2}O_{2}=4:1}$)
for 5 min, rinsed in de-ionized water and immediately loaded into
a Plassys MEB 550S evaporator. After one night of pumping, we evaporated
150 nm of Al onto the chip. Optical resist (AZ1505) was spun on the
sample and large features $(>\SI{1}{\micro\meter})$ were patterned
with UV laser lithography. After development, the wafer was etched
with Aluminium etchant, followed by cleaning in NMP overnight. We
then spun a bilayer of methacrylic acid/ methyl methacrylate (EL7),
evaporated 60 nm of Ge onto the chip and spun a high contrast electron-beam
resist (CSAR 62) on the top of the germanium layer. The qubits were
patterned by electron-beam lithography (50 kV, $\SI{660}{\micro\coulomb\per\centi\meter\squared}$).
The development took place in a 1:3 methyl isobutyl ketone (MIBK)/
isopropanol (IPA) solution for $\SI{240}{\second}$, followed by 60s
in IPA. The chip was then loaded into a Reactive Ion Etcher to perform
plasma etching with SF6 in order to form a rigid germanium mask. We
then developed the bilayer beneath with 1:3 MIBK/IPA solution for
90 s, followed by 60 s in IPA and cleaned the open regions by oxygen
ashing for $\SI{240}{\second}$. The sample was then loaded into a
Plassys MEB 550S electron-beam evaporator and pumped overnight. We
cooled the evaporator plate down to $\SI{-50}{\celsius}$ and evaporated
a first layer of 25 nm of aluminium. We then performed a dynamic oxidation
of $\mathrm{O_{2}/Ar}$ (15\%-85\%) at $P=20\:\mathrm{\mu bar}$ for
30 minutes. A second layer of 30 nm of aluminum was then evaporated
at a temperature of $\sim\SI{7}{\celsius}$ followed by a static oxidation
at $P=10\,\mathrm{mbar}$ for 10 minutes. This last step encapsulates
the junctions with aluminium oxide and allows for a more controlled
aging. An histogram of the junction resistances can be found in \ref{fig:Room-temperature-resistance}.

\begin{figure}[h]
\begin{centering}
\includegraphics[width=0.8\linewidth]{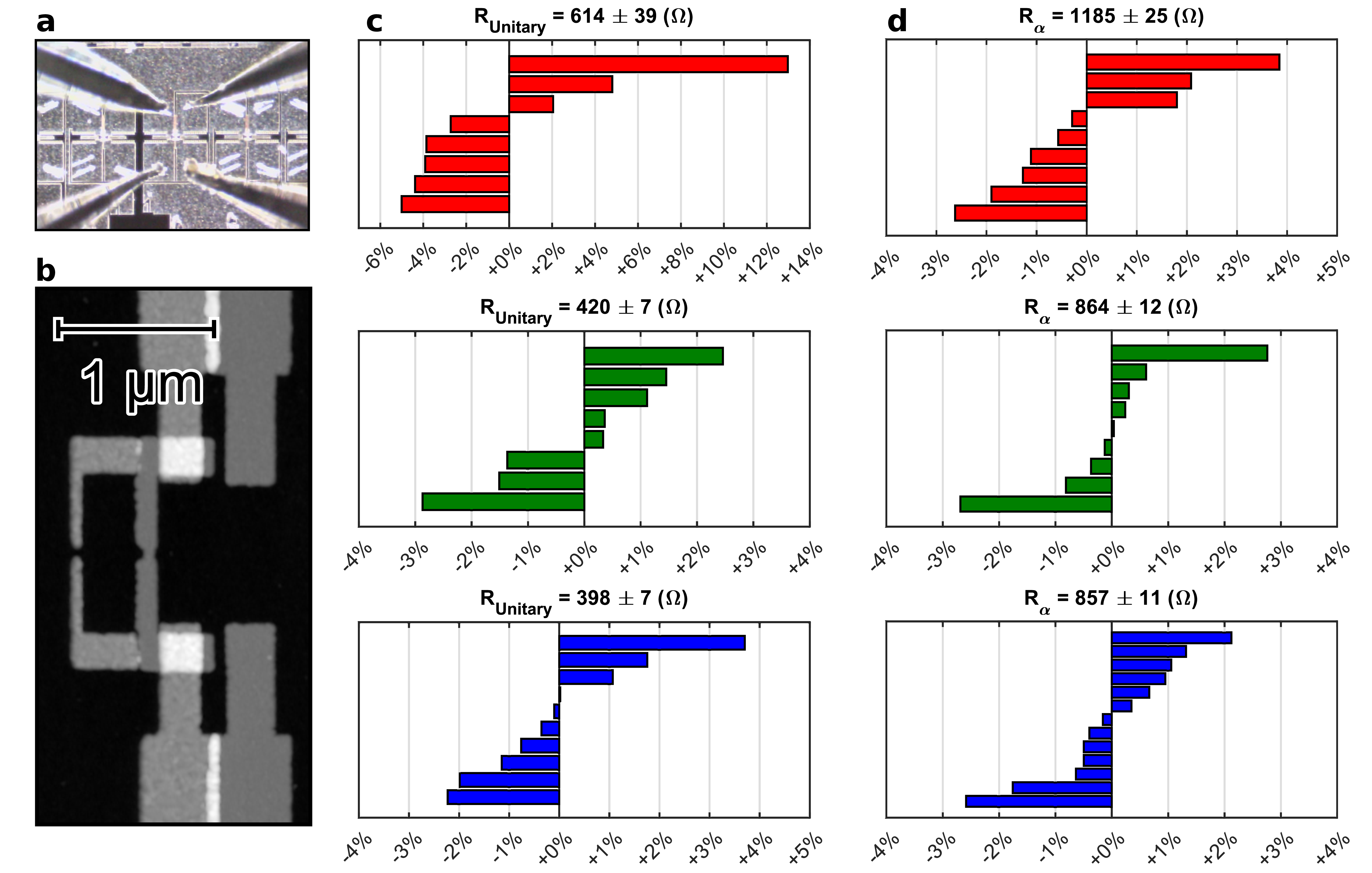}
\par\end{centering}
\caption{\textbf{Room temperature resistance measurements.} \textbf{a}, Microscope
image showing 4 probe measurement of a test sample. \textbf{b}, AFM
micrograph showing a close-up view on the test sample which consists
of the two Josephson junctions in series. \textbf{c}, Resistance measurements
of several unitary junctions for sample A (in red), B (in green) and
C (in blue). The resistance of the unitary junctions is $614\pm39\:\mathrm{\mathrm{\Omega}}$,
$420\pm7\:\mathrm{\mathrm{\Omega}}$, $398\pm7\:\mathrm{\mathrm{\Omega}}$
for sample A,B and C respectively. \textbf{d}, Resistance measurements
of several $\alpha$ junctions for sample A (in red), B (in green)
and C (in blue). The resistance of the $\alpha$ junctions is $1185\pm25\:\mathrm{\mathrm{\Omega}}$,
$864\pm12\:\mathrm{\mathrm{\Omega}}$, $857\pm11\:\mathrm{\mathrm{\Omega}}$
for sample A,B and C respectively.\label{fig:Room-temperature-resistance}}
\end{figure}
\end{widetext}

\bibliographystyle{unsrt}
\phantomsection\addcontentsline{toc}{section}{\refname}\bibliography{references_reproduciblequbits}

\end{document}